\documentclass[apj,twocolumn,twocolappendix,numberedappendix]{openjournal}
\usepackage{newtxtext,newtxmath,enumitem,multirow,ctable,verbatim}
\usepackage[T1]{fontenc}
\usepackage[dvipsnames]{xcolor}
\usepackage[breaklinks,colorlinks,citecolor=blue,urlcolor=blue]{hyperref}

\newcommand{\Alf}{{Alfv\'en}}

\newcommand{\paperone}{Paper {\small I}}

\newcommand{\orcidauthor}[3]{\author{\href{http://orcid.org/#1}{#2$^{#3}$}}}

\shorttitle{CR X-rays and Radio in Perseus}
\shortauthors{Hopkins et al.}

\begin{document}

\title{\vspace{-0.8cm}The Virgo Cluster as a Case Study for Ancient Cosmic Ray Halos in Radio and X-rays\vspace{-1.5cm}}

\orcidauthor{0000-0003-3729-1684}{Philip F. Hopkins}{1*}
%\orcidauthor{0000-0001-9185-5044}{Eliot Quataert}{2}
\orcidauthor{0000-0002-1616-5649}{Emily M. Silich}{1}
\orcidauthor{0000-0002-8213-3784}{Jack Sayers}{1}
\orcidauthor{0000-0002-7484-2695}{Sam B. Ponnada}{1}
\orcidauthor{0000-0002-1159-4882}{Isabel S. Sands}{1}
\affiliation{$^{1}$Division of Physics, Mathematics, and Astronomy, California Institute of Technology, Pasadena, CA 91125, USA}
\affiliation{$^{2}$Department of Astrophysical Sciences, Princeton University, Princeton, NJ 08544, USA}

\thanks{$^*$E-mail: \href{mailto:phopkins@caltech.edu}{phopkins@caltech.edu}},

\begin{abstract}
Virgo is the nearest cool core (CC) galaxy cluster, and is archetypal of the class of X-ray weak CCs. Despite the X-ray continuum implying a large cooling rate ($\gtrsim 10\,{\rm M_{\odot}\,yr^{-1}}$), measurements of cooler gas phases, dust, and star formation show almost no cooling outside the M87 nucleus.
It is also known that M87$^{\ast}$ is a prodigious source of cosmic ray (CR) leptons observed in radio and $\gamma$-rays, and recent radio observations have shown that the M87 radio halo is low-frequency and its spectral index steepens rapidly away from the injection region+jets, consistent with an aging, low-energy CR population. 
We demonstrate that coupling X-ray and radio to newer Faraday rotation limits on the magnetic field, this implies the CRs within the CC radius ($\lesssim 40$\,kpc) must contribute significantly to the total pressure and soft X-ray emission, through a combination of Coulomb heating the gas and inverse-Compton (IC) scattering microwave background photons to $\sim$\,keV. 
We show this can explain the apparent correspondence between jet/radio power and X-ray cooling luminosity, and why the X-ray inferred cooling rate appears much larger than otherwise observed.
This would also explain why the central X-ray inferred metallicity at $\lesssim 10\,$kpc appears to be suppressed relative to the UV/optical inferred metallicity (in a wavelength-dependent manner). 
Compared to more extreme radio+X-ray sources like Perseus, we find the observations require a softer ``escaping'' CR spectrum (escaping or accelerated outside $\gtrsim 1\,$kpc from M87$^{\ast}$).
\end{abstract}

\keywords{circumgalactic medium --- galaxies: clusters --- X-rays --- cosmic rays --- galaxies: formation}

\maketitle

\section{Introduction}
\label{sec:intro}

``Cool core'' (CC) clusters have been the subject of intense study in astrophysics for their implications in galaxy formation, ICM/CGM plasma physics, and supermassive black hole (BH) or active galactic nucleus (AGN) ``feedback'' and the quenching of star formation \citep[e.g.][]{mcnamara:2007.agn.cooling.flow.review.cavity.jet.power.vs.xray.luminosity.scalings.emph.compilation}. While X-ray luminous ``strong cool cores'' (SCCs) like Perseus have received much of the attention in this field, owing to their extreme properties (e.g.\ apparent X-ray cooling luminosities in excess of $\sim 10^{45}\,{\rm erg\,s^{-1}}$), it is important to note that the overwhelming majority of CCs by number, and the vast majority of the population-and-redshift integrated X-ray ``cooling luminosity'' in the Universe, comes from much less-extreme but more common ``weak cool core'' (WCC) systems \citep{reipreich:2002.cluster.xray.luminosity.functions,peres:no.cooling.flow.mass.corr,mittal:2011.xray.cluster.scalings.includes.rcool.lcool,koens:2013.warps.cluster.xray.luminosity.functions,bohringer:2014.cluster.xray.luminosity.functions,pacaud:2016.cluster.xray.luminosity.functions,hopkins:2025.cr.ic.clusters.detailedobs}. Archetypes of the WCC category in the local Universe include clusters like Virgo, Centaurus, Ophiuchus, and groups like NGC 5044 and HGC 62. In these systems the classically-inferred X-ray cooling luminosity $L_{X,\,{\rm cool}}$ of gas with an inferred cooling time $\lesssim 7\,$Gyr is more like $\sim 10^{43}\,{\rm erg\,s^{-1}}$, and the implied ``mass deposition rate'' that should be occurring owing to the cooling is $\dot{M}_{X,\,{\rm cool}} \sim 10\,{\rm M_{\odot}\,yr^{-1}}$, interior to an apparent ``cooling radius'' $R_{X,\,{\rm cool}}$ of order $\sim 20-40\,$kpc \citep{cavagnolo:cluster.entropy.profiles,hudson:2010.cool.core.cluster.review.central.properties.temperature.entropy.coolingtime.definition.coolcore.basic.scalings.size.luminosity.mdot,mittal:2011.xray.cluster.scalings.includes.rcool.lcool,liu:2024.strong.lradio.lxray.connection.luminous.cool.cores.strong.central.xray.only.when.strong.radio.source}. 
The classical ``cooling flow'' (CF) problem in these clusters is defined by the fact that they show little to no evidence for any such cooling. Although in SCCs like Perseus there is almost always some significant multi-phase gas, in WCCs like the specific clusters above, the upper limits to the neutral, warm ionized ($\lesssim 10^{6}\,$K), molecular, and dust masses are typically $\lesssim 10^{5}-10^{7}\,{\rm M_{\odot}}$ outside of the galactic nuclei deep within the ISM of the BCG (central couple kpc; see \citealt{valluri:1996.hgc.62.halpha.very.low.just.stellar.illumination,gallagher:2008.hcg.group.dust.ir.sfr.constraints,mittal:2011.centaurus.central.cool.gas.dynamics,hoffer:2012.uv.ir.sfrs.bcgs.compilation,sparks:2012.cool.gas.virgo.m87.confined.low.mass,hamer:2012.ophiuchus.other.clusters.halpha.weak.and.offset,werner:2013.virgo.cold.gas.low.mass.low.column.confined.center,werner:2014.cold.gas.giant.ellipticals.low.mass.all.ism,alighieri:2013.clusters.early.types.dust.content.survey.in.virgo,salome:2016.centaurus.molecular.atomic.gas.low.sfe,lakhchaura:2018.cluster.giant.es.xray.lum.and.radio.and.cold.gas.estimates.halpha,boselli:2019.virgo.ionized.gas.halpha.filament.measurements.low.mass.sfr,olivares:2019.cold.gas.filament.coolcore.compilation.small.to.large.mass.small.centaurus,rose:2019.bcg.molecular.gas.nuclei.estimates,richtler:2020.fornax.cluster.bcg.dust.atomic.gas.all.from.stellar.mass.loss}). And their past Gyr-averaged star formation rates (SFRs) $\dot{M}_{\ast}$ are robustly constrained by a combination of UV, optical, and infrared spectroscopic observations plus (in cases like M87 in Virgo) additional novae and supernovae rates, (lack of) young star clusters or high-mass X-ray binaries, to be $\dot{M}_{\ast} \lesssim 0.1-0.2\,{\rm M_{\odot}\,yr^{-1}}$.\footnote{More recent analyses of the central regions of M87 have argued that nova rates \citep{matteucci:2003.m87.nova.rate.and.sfr.constraints.very.low.1em4,lessing:2024.hst.m87.nova.rate.enhanced.along.jet.but.low}, resolved CMDs \citep{cook:2020.sfhs.m87.extremely.low.recent.sfr.from.pixel.cmds.also.metallicities.of.stars}, and UV spectra \citep{tamhane:2025.m87.extremely.low.sfr.uv} place a much more stringent $\lesssim 10^{-4}\,{\rm M_{\odot}\,yr^{-1}}$ upper limit on the recent star formation rate. Meanwhile previous claims of molecular gas detections even at sub-kpc scales appear to be dominated by contamination or intervening structures not associated with M87, with upper limits on the total molecular gas mass now revised to $ \ll 10^{5}\,{\rm M_{\odot}}$, with maximum molecular surface density  $\lesssim 1\,{\rm M_{\odot}\,pc^{-2}}$ in the central $<100\,$pc \citep{kawanaka:2025.m87.nucleus.co.contaminated.very.strong.upper.limit.molecular.gas,boizelle:2025.m87.co.survey.very.strong.upper.limits.almost.no.molecular.gas}.}

There are other puzzles as well. For example, in all of the above clusters, the inferred metallicity from X-ray spectra (fit to single-temperature models) of the central $\lesssim 1-10\,$kpc appear to be shockingly low, $Z_{X}(R \ll 1-10\,{\rm kpc}) \sim 0.1-0.3\,Z_{\odot}$ \citep{bohringer:2001.m87.cluster.profiles.no.real.cooling.flow.mass.deposition,molendi:2001.m87.abundance.profiles.xray.fit.out.with.multi.temp.plus.single.temp.plus.power.law.plus.background,matsushita:2002.m87.virgo.cluster.obs.metallicity.drop.temperature.fitting.challenges,churazov:2008.virgo.mass.profile.kinematics.vs.xrays.as.measurement.of.nonthermal.pressure,russell:2015.m87.highres.chandra.zprofile.superstrongdrop,russell:2018.m87.inner.kpc.multi.temp.and.strong.agn.maps.with.zdrop.many.signatures.of.cr.ic}, but for \textit{the same radii and galaxies} there are robust measurements of both gas-phase and stellar-phase metallicities from standard UV/optical/near-IR diagnostics that all show similar modestly-super-Solar ($Z_{\rm opt/UV/NIR} \sim 1.5-2\,Z_{\odot}$) metallicities \citep[e.g.][]{olivares:2025.massive.group.cluster.core.gas.metallicity.gradients.none.show.drops.like.in.xrays.even.drops.much.less.extreme.optical.UV}.  And the X-ray inferred ``hydrostatic equilibrium'' mass profiles interior to $\lesssim 10\,$kpc appear (in more modern, more accurate measurements) to disagree by factors of several with the mass profiles from a combination of stellar and gas and globular cluster kinematics \citep{das:2010.m87.xray.hydrostatic.mass.profiles,murphy:2011.new.m87.models,agnello:2014.m87.gc.kinematics.updated.mass.models,liepold:2023.m87.dynamical.masses}.  

Moreover, the small CCs of these halos also show -- at the same radii -- low-frequency radio halos. 
In Virgo, for example, within the central $\sim 20-40\,$kpc around M87 there is diffuse, volume-filling radio emission detected by LOFAR and other instruments at frequencies as low as $20\,$MHz \citep{owen:2000.m87.radio.maps.90cm.330MHz,degasperin:2012.m87.20mhz.to.10.ghz.maps.super.spectrally.resolved.extended.halo.very.steep.steepening.further.out,edler:2023.lofar.144MHz.m87.halo.images,degasperin:2025.virgo.50.to.1500.mhz.radio.halo.images}. While clearly powered by the famous BH at the center of M87 and its associated jets, the diffuse spectrum is now seen to be (1) very steep even at $\sim 10\,$kpc from the radio core or jet termination shocks, and (2) to steepen rapidly with increasing distance from those source regions, with most of the diffuse halo volume showing an ``ultra steep'' spectrum with very clear spectral curvature. Indeed the high-frequency portion of the spectrum appears to steepen to spectral indices as steep as $\alpha \sim 2.5-4$ before becoming undetectable at $\gtrsim 10\,$kpc, where the lowest-frequency halo persists to larger radii comparable to the CC \citep{rottmann:1996.effelsberg.m87.10ghz.images.70arcsec.resolution,degasperin:2012.m87.20mhz.to.10.ghz.maps.super.spectrally.resolved.extended.halo.very.steep.steepening.further.out,mtshweni:2022.meerkat.ghz.images.spectral.indices.m87,wu:2025.m87.radio.multifrequency.maps.and.spectral.hardening.strongly.outside.core}. This requires a large population of diffuse, low-energy cosmic rays (CRs). 

In \citet{hopkins:2025.cr.ic.clusters.ideapaper}, we suggested that many puzzles in CCs like those above could be naturally explained by ``ancient'' cosmic ray halos (ACRHs) within the CC. CR leptons accelerated by the radio galaxy stream outwards, but the high-energy ($E_{\rm cr} \gg $\,GeV) CRs rapidly lose their energy, leaving a population with energy at $\sim 0.1-1\,$GeV which simultaneously explain the low-frequency radio halo and inverse-Compton (IC) scatter CMB photons to $\sim$1\,keV energies, producing a remarkably thermal-continuum-like soft X-ray spectrum \citep{hopkins:2025.crs.inverse.compton.cgm.explain.erosita.soft.xray.halos} with substantial curvature (owing to the curvature of the CR spectrum imprinted by losses). These CR-IC X-ray halos dim as they stream out but can reach $\sim 100\,$kpc before losing all their energy. 
In \citet{hopkins:2025.cr.ic.clusters.detailedobs}, we showed that these ACRHs would naturally explain a number of properties of CCs, including their characteristic ``universal'' X-ray inferred surface brightness (SB) and density/temperature/pressure/entropy profiles, and alleviate the CF problem because the X-ray continuum is boosted by CR-IC (so the true cooling luminosity can be much lower). We showed they also trivially explain the various observed correlations between AGN/radio/jet/cavity power and $L_{X,\,{\rm cool}}$, as both trace the same lepton population; as well as other properties like the CC size-luminosity relation. \citet{hopkins:2025.cr.ic.perseus}, henceforth \paperone, applied these models to a detailed case study of Perseus, as archetypal of SCCs, and showed that the expected ACRH from the BCG (NGC 1275/3C 84, with leptonic injection rate $\sim 10^{45}\,{\rm erg\,s^{-1}}$) can explain the spatially-and-spectrally observed properties from radio ($\gtrsim 10\,$MHz) through soft and hard X-ray and $\gamma$-ray ($\lesssim$\,TeV) from $R\sim 1-100\,$kpc within the apparent SCC. And \citet{hopkins:2025.cr.ic.neutral.gas.clusters} showed the associated low-energy CR population naturally explains the otherwise extremely puzzling high-excitation properties of the neutral atomic and molecular gas in the diffuse SCC (distances $\gtrsim 10\,$kpc from the BCG, with excitation more similar to gas observed in strong CR environments like pulsar wind nebulae, e.g.\ the Crab, as compared to the local interstellar medium [LISM]; \citealt{ferland:2008.molecular.emission.in.cooling.flow.filaments.shielding.excitation.models,salome:2011.extended.molecular.gas.around.perseus.cluster.core,bayet:2011.cosmic.ray.heating.needed.molecular.lines.in.perseus,canning:2016.Halpha.CII.OI.CO.modeling.excitation.line.ratios.skin.depths.extreme.excitation.cold.atomic.gas}). 
Given the broadly similar but quantitatively distinct puzzles above, associated with WCCs, it seems important to consider if the same models, scaled to the properties of a typical WCC halo and BCG like M87, can explain a similar range of phenomena. 

In this paper, we therefore apply these same ACRH models from \paperone\ (with appropriately rescaled properties) to the best-studied WCC, Virgo. 
We show that a simple model, assuming M87$^{\ast}$ accelerates $\sim 10^{43}\,{\rm erg\,s^{-1}}$ of CR leptons, can explain the apparent WCC X-ray apparent SB/density/temperature/pressure/entropy/mass deposition rate/cooling luminosity interior to the CC (while remaining below hard X-ray and $\gamma$-ray upper limits) and the radio ``mini-minihalo'' spectrum from $\sim 10-10^{3}\,$MHz and radial profile out to $\sim 40\,$kpc. It can also potentially explain the discrepancy between X-ray and NIR/optical/UV-inferred metallicities at $\sim1\,$kpc. 
\S~\ref{sec:model} summarizes the model and changed inputs from \paperone. \S~\ref{sec:results} shows the resulting predictions for X-ray properties/profiles, metallicities, the central radio profile, and high energy constraints, as well as SZ and kinematic/mass/potential models. \S~\ref{sec:conclusions} summarizes and discusses these results.

\begin{figure}
	\centering
	\includegraphics[width=0.98\columnwidth]{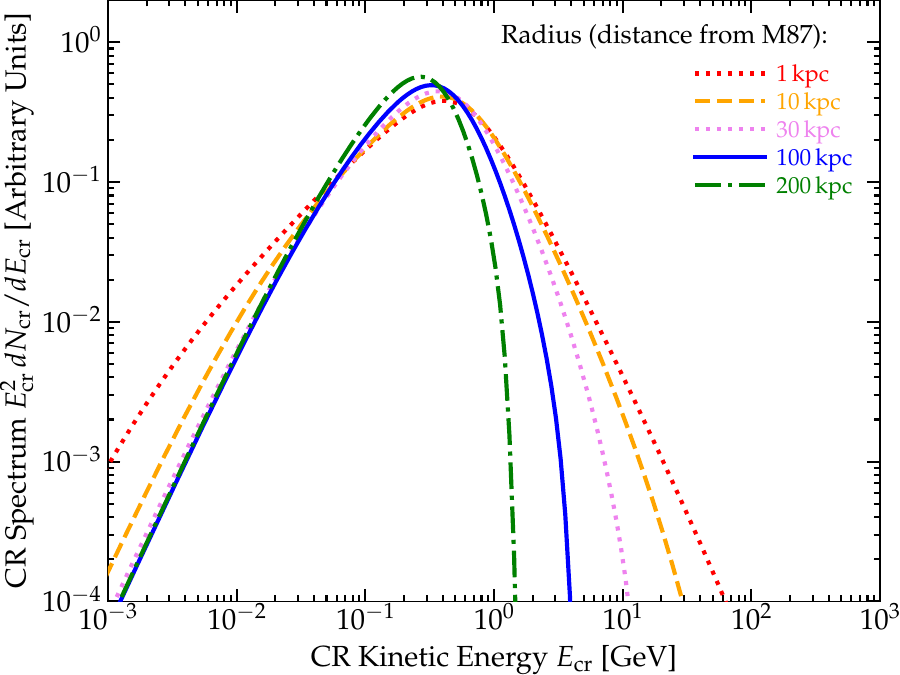}
	\caption{Predicted CR spectra (\S~\ref{sec:cr.spectra}; arbitrarily normalized to emphasize spectral shapes) at different cluster-centric radii $r$, for the fiducial/example Virgo model we consider (\S~\ref{sec:model}), of CRs injected at the center of M87 streaming outwards. Losses (synchrotron+IC at high-$E_{\rm cr}$, Coulomb at low-$E_{\rm cr}$) produce substantial spectral curvature by the time CRs travel $\gtrsim$\,kpc distances, with a clear peak at $\sim 0.1-1\,$GeV where losses are minimized.
	\label{fig:crspec}}
\end{figure}

\section{Model}
\label{sec:model}

We consider a simple analytic model defined in \paperone. Specifically, we adopt a spherically symmetric cluster with a specified diffuse/volume-filling gas density $n$, temperature $T$, metallicity $Z$, mass $M(<r)$, and magnetic field $B$ profile, with CRs obeying the steady-state, spherically-symmetric Fokker Planck equation: 
\begin{align}
\label{eqn:onemoment} 0 &=  \frac{1}{r^{2}} \frac{\partial }{\partial r}  \left[ r^{2} \left( \kappa \,\frac{ \partial f}{\partial r} - v_{\rm st} f \right)\right]
 + \frac{\partial (p \mathcal{R} f ) }{\partial p} + S(r,p) \ .
\end{align}
in terms of an effective diffusivity $\kappa$, streaming speed  $v_{\rm st}$ (which includes \Alf{ic} or diffusive-like ``super-\Alf{ic}'' streaming, or super-diffusion, or advection/convective transport), losses $\mathcal{R}$, and source term $S$. The source term is treated as a point-like injection in a sphere at $R<1\,$kpc -- what we refer to here as the ``injection zone'': we only evolve the CRs outside of $>1\,$kpc, since we are only interested in the diffuse gas and CRs, not the inner AGN/jets/lobes/etc. or acceleration mechanisms. Thus what matters is the spectrum \textit{emerging} from these radii. 
For background properties we adopt simple fits to the properties at large radii and constraints from non X-ray observations, extrapolated to the central regions as if for a non-cool-core cluster: 
$(n/{\rm cm^{-3}})^{2} = n^{2}_{0}\,(r/r_{c})^{-a_{0}}\,\{ [1 + (r/r_{c})^{2}]^{(3\,a_{1}-a_{0})/2} [1 + (r/r_{s})^{3}]^{a_{2}/3} \}^{-1}$ with $(r_{c},\,r_{s},r_{500},\,a_{0},\,a_{1},\,a_{2},\,n_{0}) = (0.04\,r_{500},\,0.6\,r_{500},660\,{\rm kpc},\,0.5,\,0.35,\,2.6,\,0.005)$; 
$k_{B} T/{\rm keV} = 3.6\,\{ 1 + (r/0.7\,r_{500})^{2}  + (r/r_{500})^{10} \}^{-0.3}$; 
$Z/Z_{\odot} = 0.1 + [0.5 + r/r_{c}]^{-1}$; 
$(B/{\rm \mu G})^{-2} = [0.1 + 50\,{\rm kpc}/r]^{-2} + 15^{-2}$, 
and $M(<r)$ corresponding to the sum of a central isothermal sphere profile with $V_{c} = 500\,{\rm km\,s^{-1}}$ plus an NFW profile with $M_{\rm vir} = 1.5 \times 10^{14}\,M_{\odot}$ and concentration $c_{\rm vir} = 7.5$. 
We stress that the details of these profiles are not important to our conclusions: all that matters is that they give us a plausible fit to larger-radii behavior, while intentionally extrapolating to values \textit{below} the X-ray inferred $n_{X}$ in the cluster center, and giving a plausible radio profile from e.g.\ $B$ and the CR models we will study (for example, considering a halo more massive by factors $\sim 5-7$, per \S~\ref{sec:other}, has no effect on our predictions). We take the diffuse radiation field at each $r$, needed to calculate IC effects, to be given by the sum of the metagalactic backgrounds+CMB compiled from \citet{cooray:2016.extragalactic.background.light.compilations.review,khaire:2019.extragalactic.background.light.spectra}, plus the observed M87 starlight+dust spectrum from far-IR through EUV, plus the compact M87$^{\ast}$ radio+$\gamma$-ray spectrum, assuming this decreases as $\propto 1/r^{2}$ (compiled from NED and \citealt{berghofer:1999.euv.virgo.extended.emission,bonamente:2001.euv.excess.emission.virgo.cluster,jk:profiles,rudick:2010.virgo.icl.colors,krick:2011.virgo.icl.ir} in \citealt{alfaro:2022.m87.broadband.sed.compilation.to.high.energy.constraints.power.injection} and shown in \S~\ref{sec:hx.gamma}), plus the emitted thermal gas-cooling (primarily X-ray) plus CR emission (radio+X-ray+$\gamma$-ray) predicted (like \paperone, this means we iterate the models, calculating the initial emission, then using this to update the radiation fields seen by CRs, re-calculating the CR emission spectrum, and so forth, until convergence). 

For the CRs we adopt a leptonic injection rate $\dot{E}_{\rm cr} = \int_{0}^{\infty} \int_{0}^{\infty} 4\pi r^{2} S\,dr dp = 3\times 10^{43}\,{\rm erg\,s^{-1}}$, motivated by direct observations and modeling of the M87$^{\ast}$ core radio and high-energy $\gamma$-ray emission spectrum and variability, which argue for leptonic injection rates in the range $\sim 4-10\times 10^{43}\,{\rm erg\,s^{-1}}$ \citep[e.g.][and references therein]{honda:2007.m87.gamma.ray.spectrum.variability.modeling.cr.injection.rate,alfaro:2022.m87.broadband.sed.compilation.to.high.energy.constraints.power.injection,hakobyan:2025.m87.core.power.accelerating.1e44.erg.s.in.crs.mostly.leptons}. Since our injection zone is the central kpc of M87, which has conditions similar to the local ISM through which the CRs must have already propagated and been processed, we assume a default leptonic injection-zone spectrum following the observed LISM lepton spectrum from \citet{bisschoff:2019.lism.cr.spectra}, as in \citet{hopkins:2025.crs.inverse.compton.cgm.explain.erosita.soft.xray.halos,hopkins:2025.cr.ic.clusters.ideapaper,hopkins:2025.cr.ic.clusters.detailedobs}. This is also consistent with the spectrum that would emerge from many processes, such as pure power-law injection with a slope $d\dot{N}/dp \propto p^{-2.2}$ after processing in the central kpc by synchrotron and inverse-Compton losses, or initially ``hadronic'' injection by pionic processes from an initially hadronic CR spectrum traveling through a dense column on sub-kpc scales (although \citealt{atoyan:2003.hadronic.blazar.jets.m87.and.other.examples} and \citealt{alfaro:2022.m87.broadband.sed.compilation.to.high.energy.constraints.power.injection} argue a hadronic jet would require an unrealistic $>10^{50}\,{\rm erg\,s^{-1}}$ injection power, and \citealt{hess:2023.claimed.upper.limit.cr.pressure.inner.virgo.from.gamma.rays.incorrect.assumptions.all.hadronic.powerlaw.no.curvature.completely.trapped.etc} place upper limits on the hadronic mass escaping into the cluster). For simplicity we adopt a constant in space-and-time $\kappa = 10^{28}\,{\rm cm^{2}\,s^{-1}}\,f(E_{\rm cr})$ and $v_{\rm st} = 100\,{\rm km\,s^{-1}}\,f(E_{\rm cr})$, with $f(E_{\rm cr}) \sim 1 + (E_{\rm cr}/{\rm GeV})^{1/2}$ consistent with standard fits from codes like GALPROP/DRAGON2/PICARD/USINE as well as CR-MHD simulations in the LISM \citep{
weinrich:2020.usine.halo.size.modeling,zhao:2021.spatially.dependent.cr.propagation.disk.halo.models,hopkins:cr.multibin.mw.comparison,hopkins:2021.sc.et.models.incompatible.obs,korsmeier:2022.cr.fitting.update.ams02,dimauro:2023.cr.diff.constraints.updated.galprop.very.similar.our.models.but.lots.of.interp.re.selfconfinement.that.doesnt.mathematically.work,jacobs:2023.isotope.ratios.large.halo.size.required.constrain.volume.of.inhomogenous.slower.diffusion.zones.in.model,silver:2024.cr.propagation.low.energies.new.data,delatorre.luque:2024.gas.models.of.galaxy.key.for.scale.height.but.need.halo.for.crs,tovar:2024.inhomogeneous.diffusion.cr.spectra,ramirez:2024.3d.struct.galaxy.gas.influences.cr.fitting.params.vs.2d,recchia:2024.cr.spectral.modeling.features.bumps.wiggles}.
Like in \paperone, we have considered variations to all of the above choices. Some have almost no effect on our most of our predictions (e.g.\ $M(<r)$, or $n(r)$ at small radii, which is only important for the very-central metallicity profile predicted, shown in \S~\ref{sec:soft.xr}). Some, like $B$, are allowed to have a wide range of values (shown in \S~\ref{sec:B}), with the default choice here chosen in all cases to be the most ``conservative'' (the one for which it is least-trivial to explain X-ray properties via CR-IC), but these are effectively degenerate with other choices like the shape of the injection spectrum for some properties (like the radio spectrum in the central $\lesssim 10\,$kpc). The shape of the injection spectrum only matters very close to the injection zone, as shown in \citet{hopkins:2025.cr.ic.perseus,hopkins:2025.cr.ic.clusters.ideapaper,hopkins:2025.cr.ic.clusters.detailedobs,hopkins:2025.crs.inverse.compton.cgm.explain.erosita.soft.xray.halos,hopkins:2025.cr.ic.neutral.gas.clusters} and \paperone, as the combination of Coulomb+synchrotron+IC losses very quickly force it to converge to a universal shape, so one should treat the predictions at the smallest radii as more sensitive to these choices, while the predictions at $R \gtrsim 10\,$kpc are more robust. The total injection luminosity of course just scales the predictions up/down accordingly. There is also some degeneracy between the values of $\kappa$ and $v_{\rm st}$, and whether we assume these are energy-dependent or not, as in the LISM, and we discuss this in \S~\ref{sec:sccs}. 

The important point is that this is a simple choice of a ``fiducial'' model, and as should be obvious many of the values we adopt for parameters are simply an approximate ``round number'' -- these are not in any way fitted to the observations. In fact in exploring the parameter space we find other model choices which slightly better-fit some of the observational datasets to which we compare (but we stay with the fiducial values as those shown, because of their simplicity). But like in \paperone, there are many model variants that fit the observations comparably well, especially owing to partial parameter degeneracies. The variations and their effects are shown therein -- here we choose to show just a single fiducial model for clarity, because it illustrates the key point, that a simple model for CR-IC can plausibly explain a wealth of multi-wavelength observations.

\section{Results}
\label{sec:results}

The key predictions from these models, and their comparison to observations, are presented here, at wavelengths from $10^{6}-10^{26}$\,Hz, and cluster-centric radii $\sim 1-1000\,$kpc.

\begin{figure*}
	\centering
	\includegraphics[width=0.32\textwidth]{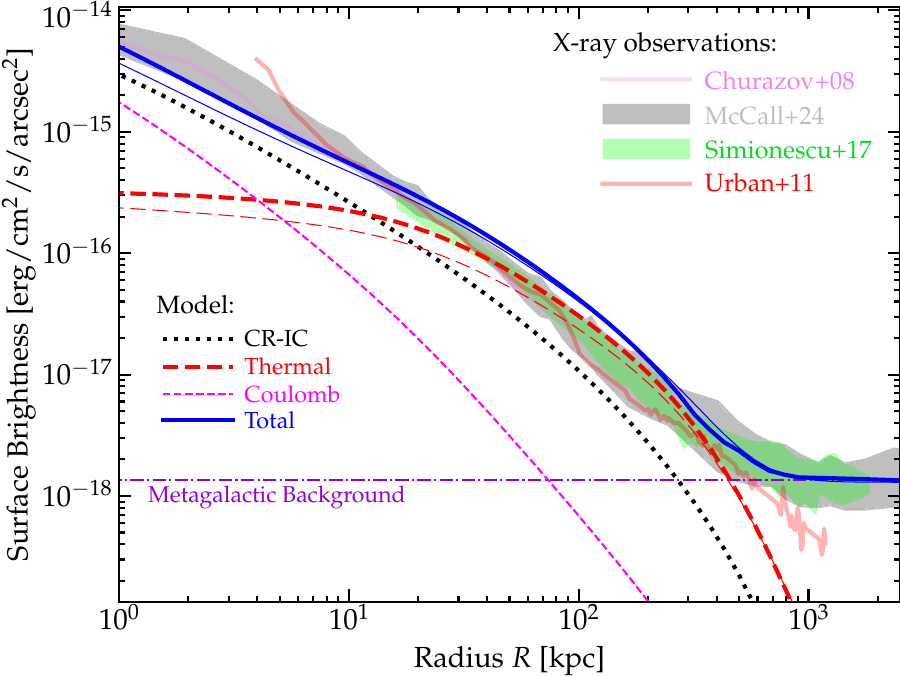}
	\includegraphics[width=0.32\textwidth]{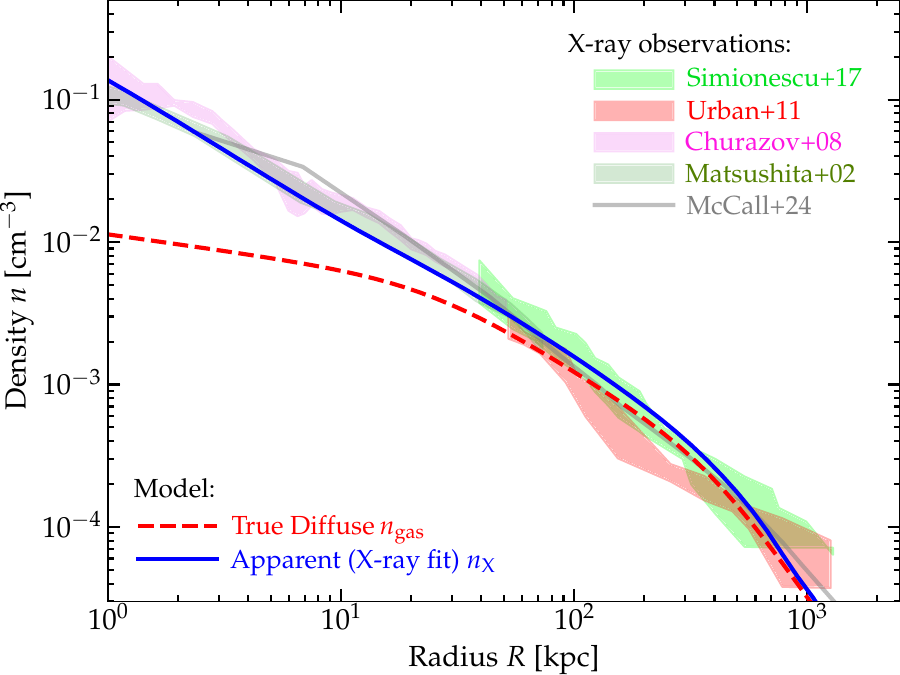}
	\includegraphics[width=0.32\textwidth]{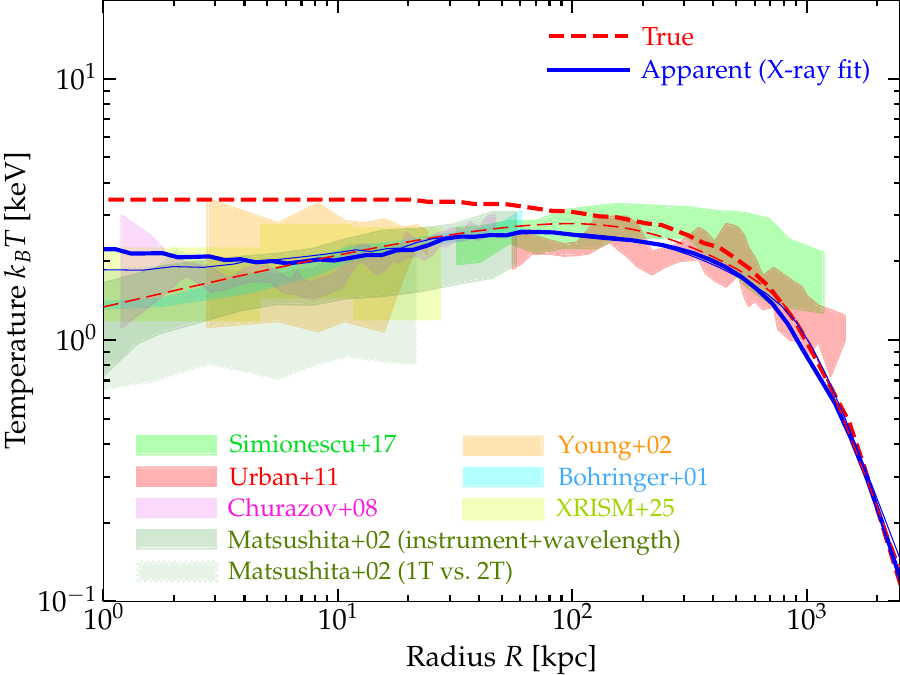}
	\includegraphics[width=0.32\textwidth]{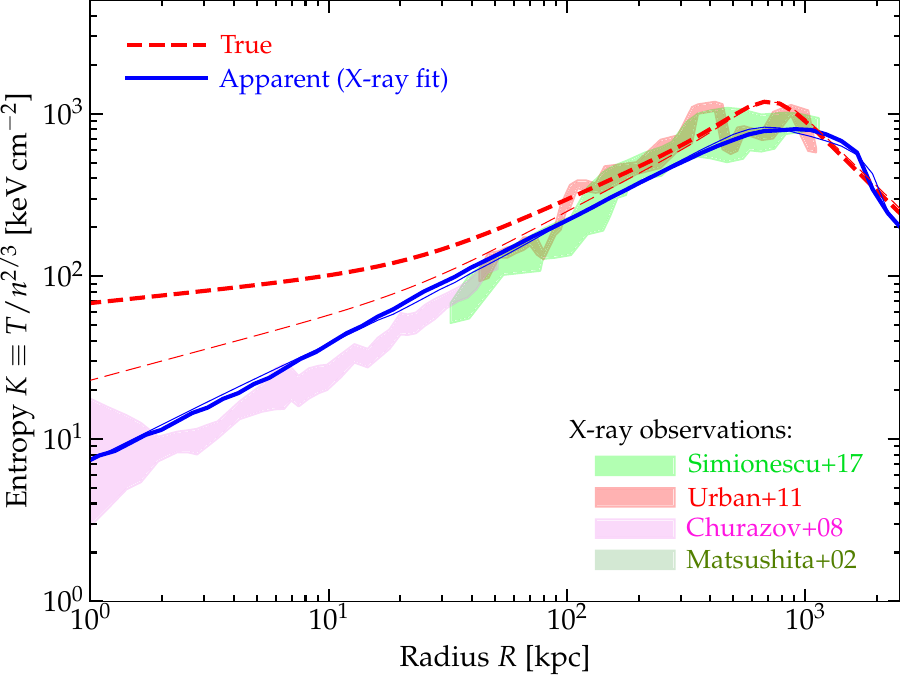}
	\includegraphics[width=0.32\textwidth]{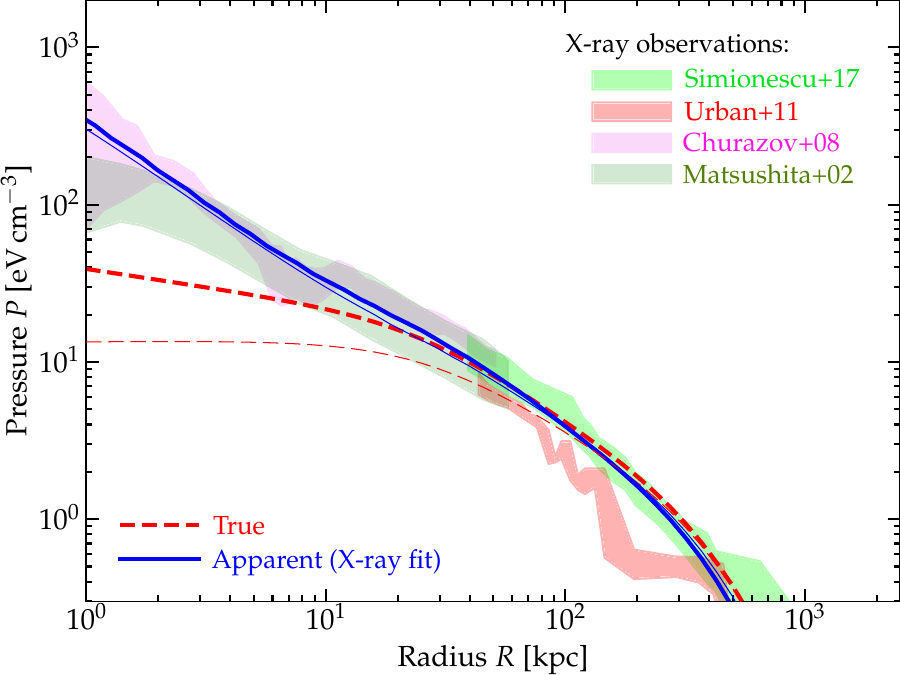}
	\includegraphics[width=0.32\textwidth]{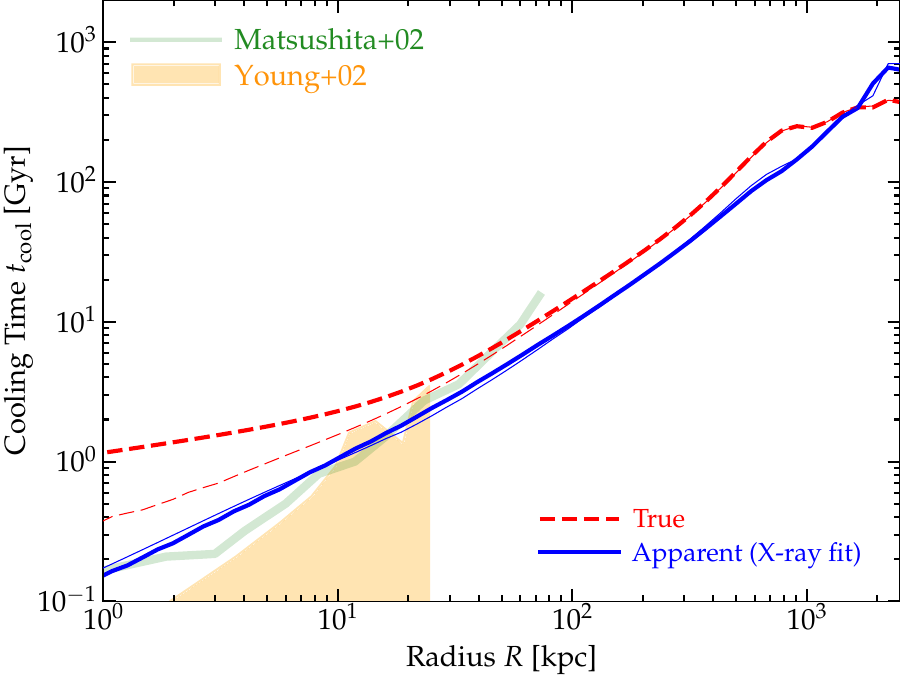}
	\includegraphics[width=0.32\textwidth]{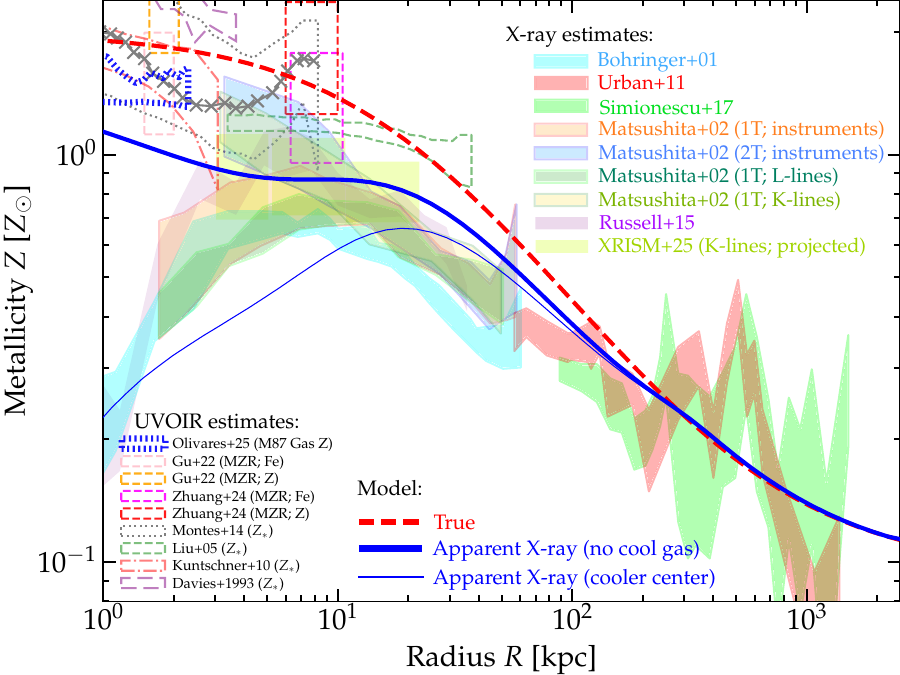}
	\includegraphics[width=0.32\textwidth]{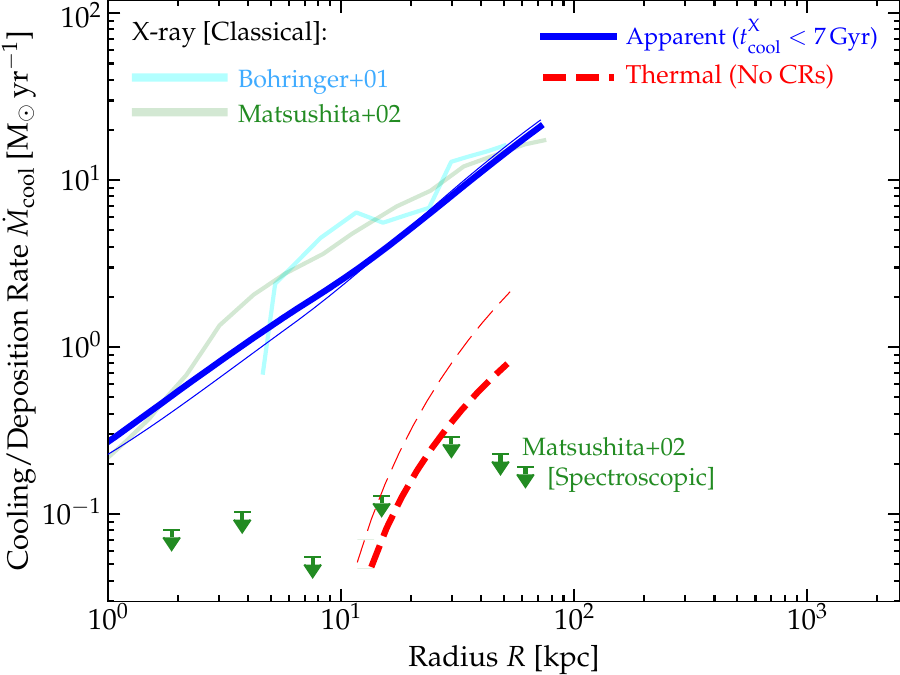}
	\caption{Predicted-vs.-observed soft X-ray properties of Virgo (\S~\ref{sec:soft.xr}). We show surface brightness (projected), and de-projected density; temperature; entropy; pressure; cooling time; metallicity; cooling ``mass deposition rate'' (for gas with $t_{\rm cool} < 7\,$Gyr). For each we plot the ``true'' underlying thermal property assumed, and the apparent X-ray inferred value accounting for CRs. We compile X-ray observations as shown: where available, shaded ranges show the interquartile range observed at each radius. CR-IC could account for a significant fraction of the apparent CF in the Virgo center, reproducing many of the observed trends with a ``true'' CF as weak as $\sim 0.1\,{\rm M_{\odot}\,yr^{-1}}$. Thin lines show an alternative model with a different gas temperature (CRs supporting cooler gas in the core); this makes little difference except to $Z_{X}$, which is very sensitive to the assumed background thermal properties and modeling procedure (\S~\ref{sec:zmet}), but CR-IC models can explain the central $Z_{X}$ suppression relative to UV/optical/IR $Z$ measurements. 
	\label{fig:xr}}
\end{figure*}

\subsection{Cosmic Ray Spectra}
\label{sec:cr.spectra}

Fig.~\ref{fig:crspec} shows CR spectra at different spherical radii $r$ in mock cluster, evolved as above. As in our previous ACRH studies \citep{hopkins:2025.cr.ic.perseus,hopkins:2025.cr.ic.clusters.ideapaper,hopkins:2025.cr.ic.clusters.detailedobs,hopkins:2025.crs.inverse.compton.cgm.explain.erosita.soft.xray.halos}, the CR spectra exhibit a high degree of curvature, and converge rapidly outside of $\sim 1-10\,$kpc to the same form as they are shaped by a combination of Coulomb losses (rapid at low energies) and synchrotron+IC losses (rapid at high energies). Critically, over any non-negligible baseline in energy, they are poorly approximated by power-law spectra $dN_{\rm cr} / dE_{\rm cr} \propto E_{\rm cr}^{-\delta}$ , and especially poorly approximated by the common assumption $\delta \sim 2.2$ often invoked in searches for IC emission. As shown in those papers above, for CR spectra peaked, like these, at $\sim 0.1-1\,$GeV, there are three peaks in the direct CR emission: $\sim$\,MHz radio synchrotron, $\sim$\,keV soft X-ray IC, and $\sim 10-100\,$MeV $\gamma$-ray bremsstrahlung.

\subsection{Soft X-ray}
\label{sec:soft.xr}

%Fig.~\ref{fig:xr} shows the resulting soft X-ray properties, including all emission (CR direct emission + thermal [accounting for CR heating as \S~\ref{sec:methods}]). We show the soft X-ray surface brightness (projected) and the (deprojected) X-ray inferred density $n_{X}$ and temperature $T_{X}$, defined as \paperone\ and various X-ray observational studies \citep[e.g.][]{mazzotta:2004.xray.temperature.measurement.modeling.and.caveats} as the density and temperature from a single-temperature spectrum that give the same emissivity and mean photon index/hardness ratio around the peak of the SED, as a function of radius $R$ from the center of M87. We then show various derived quantities including entropy $K_{X}$, pressure $P_{X}$, cooling time $t_{{\rm cool},\,X}$, metallicity $Z_{X}$ (defined as in \paperone, but see e.g.\ \citealt{matsushita:2002.m87.virgo.cluster.obs.metallicity.drop.temperature.fitting.challenges} and \citealt{hopkins:2025.cr.ic.clusters.detailedobs} for discussion of the very large systematic uncertainties in both the observations and model predictions), cooling luminosity $L_{X,\,{\rm cool}}$ and classical cooling mass deposition rate  $\dot{M}_{X}$. 

Fig.~\ref{fig:xr} shows the soft X-ray properties versus radius from the model. We show the (projected 2D) surface brightness in a corrected $0.1-10\,$keV band, and the de-projected (3D) profile of the X-ray inferred density $n_{X}$, temperature $T_{X}$, and metallicity $Z_{X}$ which reproduce the same emissivity, spectral shape/hardness, and Fe K line equivalent width, as the modeled CR-IC + thermal APEC spectrum at each radius (see \paperone\ for details). We also present various derived X-ray properties including entropy $K_{X} \equiv k_{B} T_{X} / n_{X}^{2/3}$, pressure $P_{X} \equiv n_{X} k_{B} T_{X}$, cooling time $t_{{\rm cool},\,X} \equiv (3/2)\,n_{X}\,k_{B}\,T_{X} / n_{X}^{2} \Lambda(T_{X},\,Z_{X})$, and cooling or mass deposition rate $\dot{M}_{X} \equiv \int (2\mu m_{p} / 5 k_{B} T_{X}) d L_{{\rm cool},\,X}(r)$. For each, we also show the ``true'' (thermal-only) background properties assumed in the model. We compile and compare X-ray observations from \citet{bohringer:2001.m87.cluster.profiles.no.real.cooling.flow.mass.deposition,matsushita:2002.m87.virgo.cluster.obs.metallicity.drop.temperature.fitting.challenges,young:2002.chandra.virgo.core.profiles.vs.quadrants,churazov:2008.virgo.mass.profile.kinematics.vs.xrays.as.measurement.of.nonthermal.pressure,urban:2011.virgo.xray.profiles.to.rvir,russell:2015.m87.highres.chandra.zprofile.superstrongdrop,russell:2018.m87.inner.kpc.multi.temp.and.strong.agn.maps.with.zdrop.many.signatures.of.cr.ic,simionescu:2017.virgo.profiles.outskirts,mccall:2024.erosita.virgo.profiles.xr.sb.along.octants,xrism:2025.m87.xray.temperature.metallicity.velocity.central.regions.projected.estimate}, from XMM, Chandra, Suzaku, eROSITA, and XRISM. Where possible, because we are interested in the diffuse emission, we show the interquartile range along different sightlines/azimuthal angles/pixels at the same $r$ observed. 

We stress that the models here are not fit to the observations: we simply tried a few parameters as described in \S~\ref{sec:model} and show a plausible example. And they are extremely simple (spherically-symmetric, point-like LISM-like injection, time-steady, constant-$v_{\rm st}$ and $\kappa$, etc.), and as such actually involve significantly fewer parameters than the commonly-adopted fitting functions from e.g.\ \citet{vikhlinin:2006.single.multi.temperature.cluster.fits.also.includes.some.metallicity.discussion} for $n_{X}$, $T_{X}$, $P_{X}$, $K_{X}$, etc. Thus the resemblance to the observations is highly non-trivial.

\subsubsection{Alleviating the Cooling Flow Problem \&\ Explaining the AGN-Halo Connection}
\label{sec:cf}

The model in Fig.~\ref{fig:xr} provides an immediate explanation for the classical-versus-spectral cooling flow problem, defined by the apparent classical cooling rate from the X-ray continuum being at least two orders of magnitude larger than the upper limits on actual cooling from a combination of X-ray spectroscopy, infrared dust surveys, HI surveys, radio molecular gas surveys, UV star formation diagnostics, and more (see \S~\ref{sec:intro}). Because CR-IC boosts the apparent X-ray SB and therefore $n_{X}$, it boosts $L_{X}$ while decreasing $t_{\rm cool}$, greatly increasing the apparent $\dot{M}_{X}$ below some cooling time threshold (here the usual $7$\,Gyr adopted), as compared to the true cooling rate of the gas. Per \citet{hopkins:2025.cr.ic.clusters.ideapaper,hopkins:2025.cr.ic.clusters.detailedobs}, this also provides an almost-trivial explanation for the apparent close relationship between the apparent X-ray cooling luminosity $L_{{\rm cool},\,X} \sim 10^{43}\,{\rm erg\,s^{-1}}$, and AGN radio/jet power (roughly the same) -- they are powered by the same lepton source.

These are shown in a population and demographics sense in \citet{hopkins:2025.cr.ic.clusters.detailedobs}. In \paperone\ and \citet{hopkins:2025.cr.ic.neutral.gas.clusters} we consider Perseus, and argue the same can explain the difference between the apparent SCC with $\dot{M}_{X} \sim 1000\,{\rm M_{\odot}\,yr^{-1}}$ and a ``true'' underlying WCC with $\dot{M}_{\rm true} \sim 10\,{\rm M_{\odot}\,yr^{-1}}$. Here, the apparent WCC could have an even much weaker ``true'' WCC with $\dot{M}_{\rm true} \lesssim 0.1\,{\rm M_{\rm odot}\,yr^{-1}}$, but is also consistent with effectively no CC whatsoever in the underlying gas. This is outside $>1\,$kpc -- at smaller radii well within the ISM (not the CGM/halo) there is a small amount of gas and very small amount of star formation observed (\S~\ref{sec:intro}), powering M87$^{\ast}$'s accretion, but the stellar wind mass loss rate in the central sub-kpc region of M87 is expected to be $\sim 0.1\,{\rm M_{\odot}\,yr^{-1}}$, more than sufficient to power this, and there is no need (nor is there any evidence) for this to be connected to the extended ICM/CGM/halo.

\begin{figure*}
	\centering
	\includegraphics[width=0.48\textwidth]{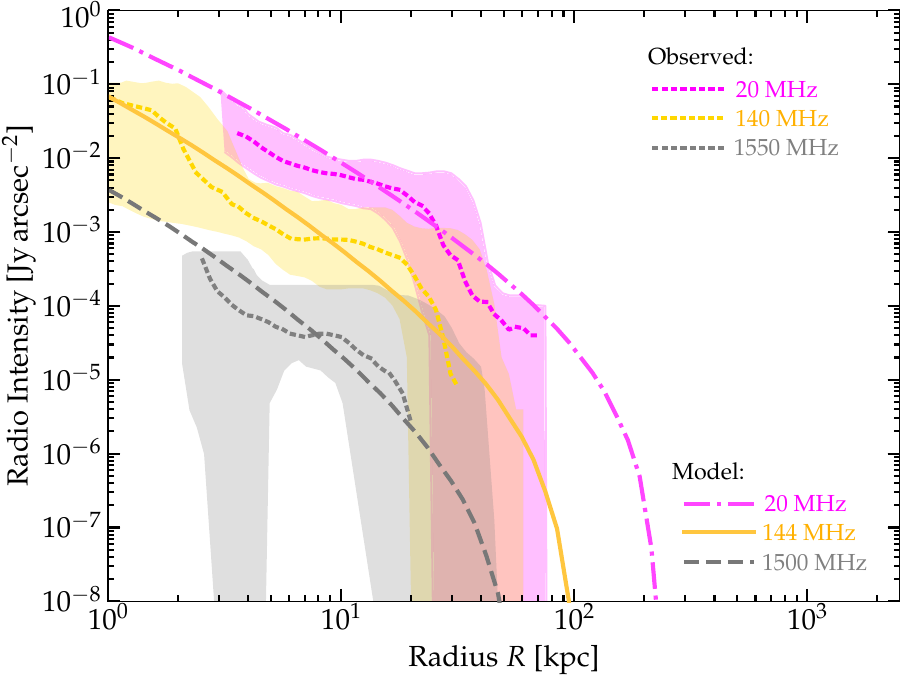}
	\includegraphics[width=0.48\textwidth]{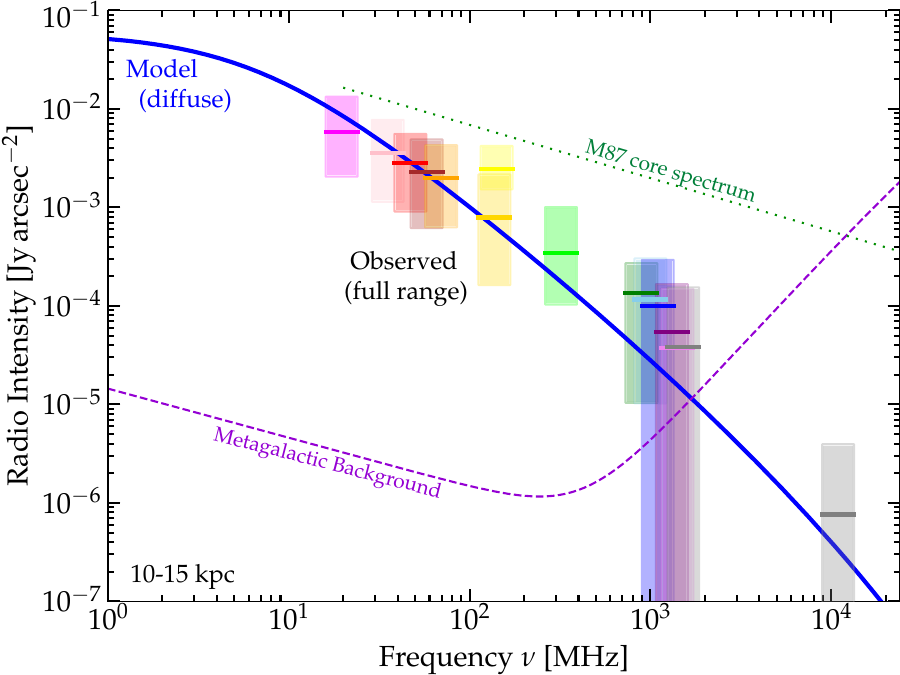}
	\caption{Predicted diffuse-gas radio emission in the Virgo/M87 halo (\S~\ref{sec:radio}).
	\textit{Left:} Intensity of diffuse synchrotron at different frequencies, from the same model as Fig.~\ref{fig:xr}. 
	\textit{Right:} Diffuse synchrotron spectrum from $10^{7}-10^{11}$\,Hz at a distance $10-15\,$kpc where this is detected over the largest frequency range (at smaller/larger-$R$, many frequencies become resolution/surface-brightness limited). 
	We compile observations from \citet{rottmann:1996.effelsberg.m87.10ghz.images.70arcsec.resolution,owen:2000.m87.radio.maps.90cm.330MHz,degasperin:2012.m87.20mhz.to.10.ghz.maps.super.spectrally.resolved.extended.halo.very.steep.steepening.further.out,mtshweni:2022.meerkat.ghz.images.spectral.indices.m87,edler:2023.lofar.144MHz.m87.halo.images,wu:2025.m87.radio.multifrequency.maps.and.spectral.hardening.strongly.outside.core,degasperin:2025.virgo.50.to.1500.mhz.radio.halo.images}; 
	and plot the median $S_{\nu}$ at each $R$ and $90\%$ inclusion interval (\textit{shaded}). 
	We contrast the (renormalized) observed M87 core-region ($\lesssim 1\,$kpc) spectrum measured, to show the clear spectral steepening.
	CR streaming+losses leads to a power-law-like profile with a frequency-dependent outer cutoff, so the overall spectrum steepens with distance from M87.
	The ``shelf'' or ``bumps'' in $S_{\nu}(r)$ and higher values at high-$\nu$ in the spectrum are well-known and spatially localized to the ``ear'' or jet termination/edge features in the 2D maps, they do not reflect the volume-filling halo, which is smoother and steeper, in good agreement with the model spectra here. Both steady spectral steepening with $R$ and significant spectral curvature at each $R$ in the diffuse gas are observed. The radio halo and spectrum are consistent with CR-IC contributing significantly to the apparent X-ray CC at these radii.
	\label{fig:radio}}
\end{figure*}

\subsubsection{Central Metallicity Suppression}
\label{sec:zmet}

As discussed in \citet{hopkins:2025.cr.ic.clusters.ideapaper,hopkins:2025.cr.ic.clusters.detailedobs}, this also can explain the apparent metallicity suppression in the Virgo center. Almost all single-temperature X-ray fits to M87 \citep{bohringer:2001.m87.cluster.profiles.no.real.cooling.flow.mass.deposition,molendi:2001.m87.abundance.profiles.xray.fit.out.with.multi.temp.plus.single.temp.plus.power.law.plus.background,matsushita:2002.m87.virgo.cluster.obs.metallicity.drop.temperature.fitting.challenges,churazov:2008.virgo.mass.profile.kinematics.vs.xrays.as.measurement.of.nonthermal.pressure,russell:2015.m87.highres.chandra.zprofile.superstrongdrop} have found that the central $Z_{X}$ fit to the X-ray spectrum either rises more slowly compared to the UV/optical metallicity or even stops rising and turns over, decreasing towards $R\rightarrow 0$. Similar results are found in most well-resolved CCs (almost all with the central $\lesssim 10\,$kpc resolved; see \citealt{matsushita:2002.m87.virgo.cluster.obs.metallicity.drop.temperature.fitting.challenges,rasmussen.ponman:2007.metallicity.temperature.profiles.groups,komiyama:2009.5044.metallicity.profile,million:2010.ophiuchus.central.metallicity.profile,murakami:2011.fornax.metallicity.profile.suzaku.xmm,walker:2013.centaurus.cluster.few.other.large.radii.entropy.metallicity.profiles,panagoulia:2015.cluster.metal.profiles.with.drops,sanders:2016.centaurus.metallicity.profiles.zdrop,gatuzz:2023.centaurus.metal.profile.outer,gatuzz:2023.ophiuchus.metallicity.profile}), including e.g.\ Virgo, Centaurus, Ophiuchus, NGC 383/2300/4325/5044/6338, HGC 42/62, Abell 262/1991/3581, and others. \citet{matsushita:2002.m87.virgo.cluster.obs.metallicity.drop.temperature.fitting.challenges} conducted a survey of systematics in M87 with XMM. They showed that the effect is weaker if one fits {only} the Fe-K line region in Virgo ($\gtrsim 4\,$keV; consistent with the recent spatially un-resolved XRISM spectra, \citealt{xrism:2025.m87.xray.temperature.metallicity.velocity.central.regions.projected.estimate}), as compared to the overall spectral shape of the X-rays ($0.5-8$\,keV) or just the Fe-L complex ($\sim1\,$keV). Similarly they found a weaker effect fitting a two-temperature model (which allows a different spectral shape and Fe-L complex behavior from Fe-K). 
\citet{molendi:2001.m87.abundance.profiles.xray.fit.out.with.multi.temp.plus.single.temp.plus.power.law.plus.background} attempted to marginalize away the $Z$-drop and found this demanded adding a range of additional temperature components \textit{and} an explicit non-thermal component contaminating the lower-energy X-rays (and their X-ray best-fit still featured highly sub-solar O and Mg with $Z_{\rm O,\,Mg} \sim 0.15-0.4\,Z_{\odot}$).  Similar results were found by \citet{russell:2018.m87.inner.kpc.multi.temp.and.strong.agn.maps.with.zdrop.many.signatures.of.cr.ic}, who still found a central drop (by a factor $\sim 4$ in $Z$) even after fitting a three-temperature (with effective temperatures $\sim 0.1-3\,$keV) plus explicit non-thermal AGN/nuclear spectrum model. 

These suggest that something has suppressed the equivalent width of metal lines in the central regions of Virgo, at lower energies $\sim$\,keV in particular. But even the Fe-K or two-temperature $Z_{X}$ appears suppressed relative to UV/optical diagnostics of the metallicity in Virgo at $\lesssim 10\,$kpc, which we compile from stellar metallicity measurements (each using different tracers and different instruments for their measurements) in \citet{davies:1993.elliptical.line.strength.mg2.metallicity.estimates,liu:05,kuntschner:2010.sauron.line.strength.metallicity.maps.ellipticals,montes:2014.m87.stellar.metallicity.age.gradients.and.central.values}. Note that much of the systematic offset between some of these measurements can be attributed to those which are sensitive to $\alpha$ abundances (e.g.\ \citealt{davies:1993.elliptical.line.strength.mg2.metallicity.estimates}) versus Fe \citep{liu:05}, with [$\alpha$/Fe]\,$\sim 0.3-0.4$ in M87 as in most massive ellipticals. We also show the \textit{gas-phase} ISM metallicity measurements (of $Z$ and Fe), including the $95\%$ range observed, for BCGs of the same stellar mass as Virgo (\citealt{gu:2022.massive.galaxy.central.metallicity.mzr.stellar,zhuang:2024.mzr.stellar.gas.different.species.starforming.and.non.very.extensive.compilation}), and the recently-measured gas metallicity values directly \textit{in M87 itself} from MUSE/VLT observations of high-resolution emission-line spectroscopy in optical/UV wavelengths \citep{olivares:2025.massive.group.cluster.core.gas.metallicity.gradients.none.show.drops.like.in.xrays.even.drops.much.less.extreme.optical.UV}. Note the latter include the effect of several different line calibration choices and template ionization spectra as part of their quoted error budget.  This discrepancy between X-ray and optical/UV determined metallicities, \textit{for gas at the same radius in the same galaxy}, follows what is predicted for CR-IC in the models here. 

As discussed in \citet{hopkins:2025.cr.ic.clusters.detailedobs} and \paperone, the central $Z_{X}$ predicted here is a very crude estimate, based on a simple ratio of the continuum emission around Fe-L and Fe-K (a proxy for the equivalent-width suppression of the lines). We do not model spectral fitting in detail, as it is well-known that $Z_{X}$ is quite sensitive to instrumental properties, fitting procedures, energy range, temperature model priors, and other choices \citep{matsushita:2002.m87.virgo.cluster.obs.metallicity.drop.temperature.fitting.challenges,mazzotta:2004.xray.temperature.measurement.modeling.and.caveats,ghizzardi:2021.iron.cluster.profiles.sensitivity.of.fitting.different.lines.different.metallicities,zhuhone:2023.cluster.temperature.fitting.sensitivities.simulations}, and we neglect (likely weaker) CR line excitation effects. Despite these simplifications, the predicted $Z_{X}$ is still quite sensitive to what we assume for the underlying thermal emission, as the lines are highly temperature-sensitive. This is obvious in Fig.~\ref{fig:xr}, and shown for a wider range of models in \paperone, as well as models in \citet{hopkins:2025.cr.ic.clusters.detailedobs} which allow for a more realistic distribution of temperatures. What is important is that (a) all models with strong CR-IC produce some central metallicity suppression relative to UV/optical, as observed, and (b) different detailed assumptions for the background gas and fitting assumptions applied to the combined spectrum can reproduce the range of observed $Z_{X}$ trends, including the more extreme central drops seen in some $Z_{X}$ estimates.

\subsubsection{On Central Temperatures}
\label{sec:TX}

Briefly we note that in Fig.~\ref{fig:xr}, we assume (for simplicity) that the true temperature $T_{\rm true}$ remains flat as $R\rightarrow 0$. Thus the suppression of the central $T_{X}$ owes to CR-IC at low X-ray energies. However, we stress that this is not necessary or important for the model. It is perfectly possible that the true temperature $T_{\rm true}$ does decrease in the Virgo center, \textit{especially} if the CR-IC scenario is true, since the prediction of the model (Fig.~\ref{fig:xr}) is that the central pressure becomes CR dominated at $R \lesssim 10-20\,$kpc, and it is very well-established that CR-pressure-dominated halos support and stabilize cooler CGM/ICM gas \citep{ji:fire.cr.cgm,hopkins:2020.cr.outflows.to.mpc.scales,ji:20.virial.shocks.suppressed.cr.dominated.halos,hopkins:2020.cr.transport.model.fx.galform,butsky:2020.cr.fx.thermal.instab.cgm,weber:2025.cr.thermal.instab.cgm.fx.dept.transport.like.butsky.study}, governed globally by buoyant dynamics \citep{hopkins:2020.cr.outflows.to.mpc.scales}. If that were the case, then the (already small) bias of $T_{X}$ from CR-IC would be even smaller, and if we modeled a somewhat harder CR spectrum it could even slightly bias the temperature higher. But importantly in any case the temperature bias from CR-IC is quite weak, or even non-existant. The important effect of CR-IC is boosting the SB/luminosity and therefore X-ray apparent \textit{density} $n_{X}$, which is what drives all of the other effects discussed above.

\subsection{Radio}
\label{sec:radio}

Fig.~\ref{fig:radio} shows the corresponding radio properties in the models here. We compile published, reduced maps of the diffuse radio surface brightness from $\sim$\,few to $\sim 40-100\,$kpc scales from: 
144 MHz LOFAR \citep{edler:2023.lofar.144MHz.m87.halo.images}; 
325 MHz VLA \citep{owen:2000.m87.radio.maps.90cm.330MHz}; 
99, 130, 166, 196, 227 MHz MWA and 
1063, 1711, 2563 MHz VLA \citep{wu:2025.m87.radio.multifrequency.maps.and.spectral.hardening.strongly.outside.core}; 
25, 36, 48, 59, 71, 140 MHz LOFAR LBA and 1400, 1600 MHz VLA \citep{degasperin:2012.m87.20mhz.to.10.ghz.maps.super.spectrally.resolved.extended.halo.very.steep.steepening.further.out}; 
54, 144 MHz LOFAR plus 1284 MHz MEERKAT \citep{degasperin:2025.virgo.50.to.1500.mhz.radio.halo.images}; 
900, 1000, 1120, 1230, 1340, 1440, 1550, 1660 MHz MEERKAT \citep{mtshweni:2022.meerkat.ghz.images.spectral.indices.m87}; and 
10550 MHz Effelsberg \citep{rottmann:1996.effelsberg.m87.10ghz.images.70arcsec.resolution}. From each, we can construct the spectrum and intensity $S_{\nu}$ in azimuthal annuli $R$, between radii from a few times the central beamsize (as defined by the authors) to an outer radius which we take to be that where the majority of pixels are above the detection limit quoted for each map. 
Importantly, we do not simply compare a linear average of the intensity in each annulus, but show the median and $\sim 90\%$ inclusion interval of the range of intensities within the annulus. Cursory examination of the maps reflects the well-known fact that at some radii, especially at higher frequencies, the mean is completely dominated by small (non-volume-filling) bright regions or hot spots (these typically correspond to the jet edges/shocks and radio ``ear'' features), while we are interested in and only model the diffuse volume-filling (lower-intensity) gas emission.

We see the simple models here reasonably reproduce the radial profiles in both normalization and shape, at frequencies from $\sim 20\,$MHz to $\sim 1.5$\,GHz. The detailed shape does differ, with the observations showing a sort of flat ``shelf'' behavior at $\sim 8-20\,$kpc, but this is well known and related to the extended jets/injection region (which we cannot capture in the models here, as we approximate the injection as point-like for simplicity). More important though is (1) the profiles are cut off by losses at roughly the correct radii; (2) their relative amplitudes are consistent, indicating that the radio spectrum is plausible and steepens with $R$ akin to the observations as CRs age; and (3) the normalization is plausible. In short, the same CR lepton density profiles which reproduce the X-ray properties are consistent with the radio properties out to tens of kpc. We also plot more detailed spectra of the predicted diffuse radio from $\sim1-10^{4}\,$MHz. These show the same, and demonstrate more clearly the un-ambiguous spectral hardening with increasing $R$ observed and discussed extensively in a number of radio studies  \citep{degasperin:2012.m87.20mhz.to.10.ghz.maps.super.spectrally.resolved.extended.halo.very.steep.steepening.further.out,mtshweni:2022.meerkat.ghz.images.spectral.indices.m87,wu:2025.m87.radio.multifrequency.maps.and.spectral.hardening.strongly.outside.core}. This suggests a population of aging leptons streaming from the central source, constraining both their power and effective streaming speed to be something like that adopted here. Indeed, the combination of the soft X-ray, radio, and Faraday rotation (\S~\ref{sec:B}) makes it challenging to construct models where CR-IC is negligible for X-rays in the core ($\lesssim 10-20\,$kpc).

While we assume a simple, spherically-symmetric model so cannot model X-ray or radio fluctuations in the 2D maps (or turbulence/shocks/etc.), these should of course be present. Several of the studies above (see e.g.\ \citealt{degasperin:2012.m87.20mhz.to.10.ghz.maps.super.spectrally.resolved.extended.halo.very.steep.steepening.further.out,degasperin:2025.virgo.50.to.1500.mhz.radio.halo.images,edler:2023.lofar.144MHz.m87.halo.images}, but also \citealt{forman:2007.m87.xray.strongly.positively.correlated.with.radio.emission.not.anti.correlated.even.along.jets} who approach the same from the X-ray side comparing to \citealt{owen:2000.m87.radio.maps.90cm.330MHz}) have shown that in the diffuse gas, like in most CC clusters at low frequencies (outside of the active jet/cocoon structure; see \citealt{giacintucci:2011.cluster.minihalo.fills.cool.core.morphologically.similar,giancintucci:2019.expanding.radio.cluster.minihalo.sample.no.good.corr.total.cluster.mass.or.total.cluster.xray.but.very.strong.corr.cooling.radius.xray.luminosity.consistent.with.linear.standard.correlation,bravi:2016.minihalo.luminosity.strong.corr.xray.luminosity.clusters,balboni:2024.lofar.xray.surface.brightess.corr.indiv.halos.well.corr.positive.as.expected.for.ic.same.particles.but.sublinear.because.spread.by.B.scatter,riseley:2023.cluster.radio.minihalo.correlated.xray.brightness.disturbed.re.energized.example,riseley:2024.large.minihalo.reenergized.with.merger.but.center.shows.strong.radio.xray.with.steeper.radio.hotter.gas.and.local.ir.ix.corr.as.expected.for.ic.nearly.linear}), there is an unambiguous \textit{positive} correlation between soft X-ray and low-frequency radio surface brightness, both globally (radial profiles) and at fixed radius (azimuthally). This is of course trivially expected in a CR-IC model (with scatter and systemic variations coming from CR acceleration/injection and $B$-field variations). 

\begin{figure}
	\centering
	\includegraphics[width=0.97\columnwidth]{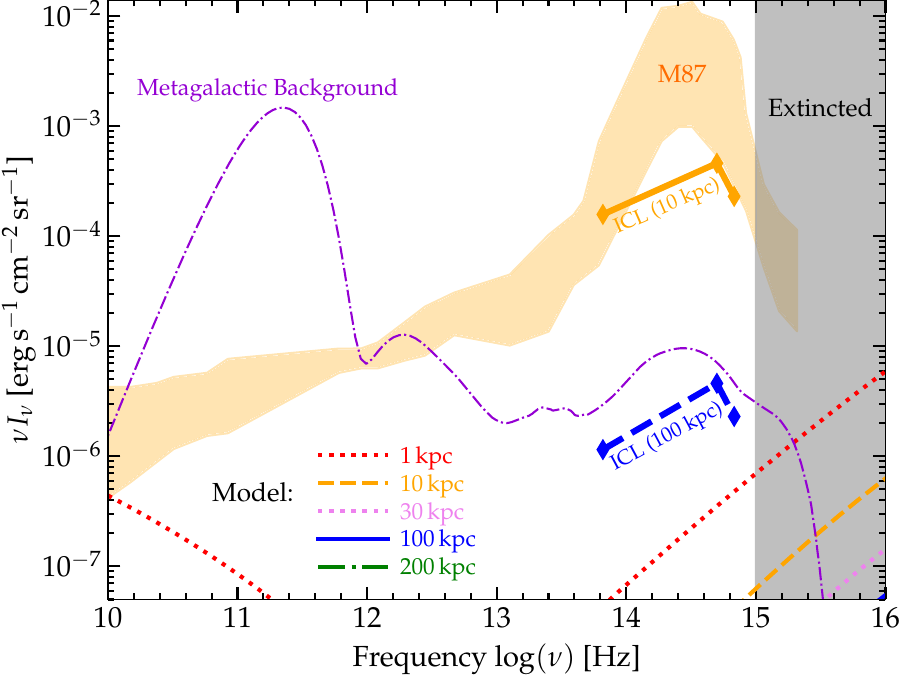}
	\includegraphics[width=0.97\columnwidth]{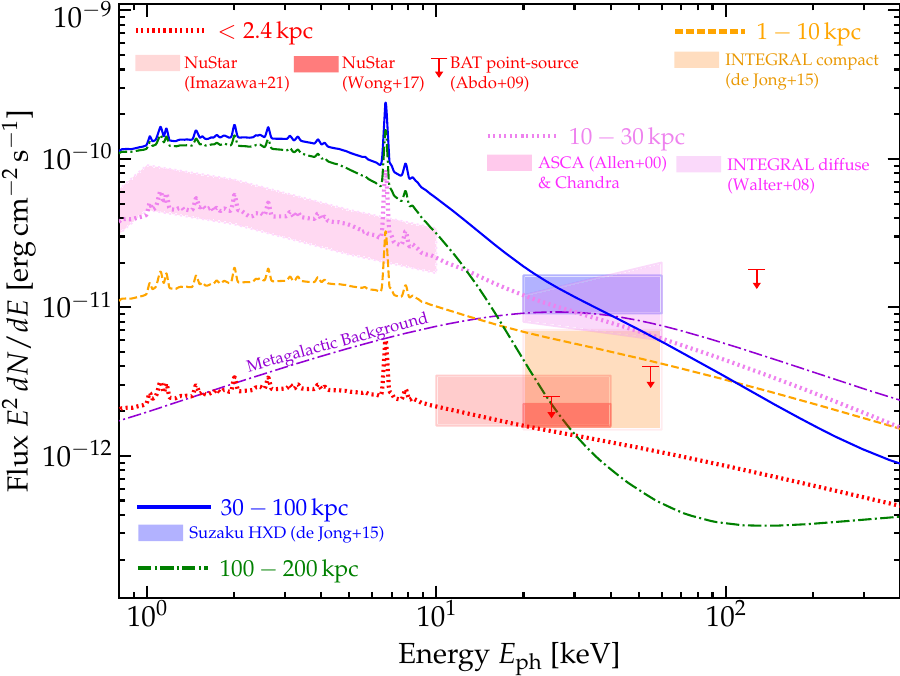}
	\includegraphics[width=0.97\columnwidth]{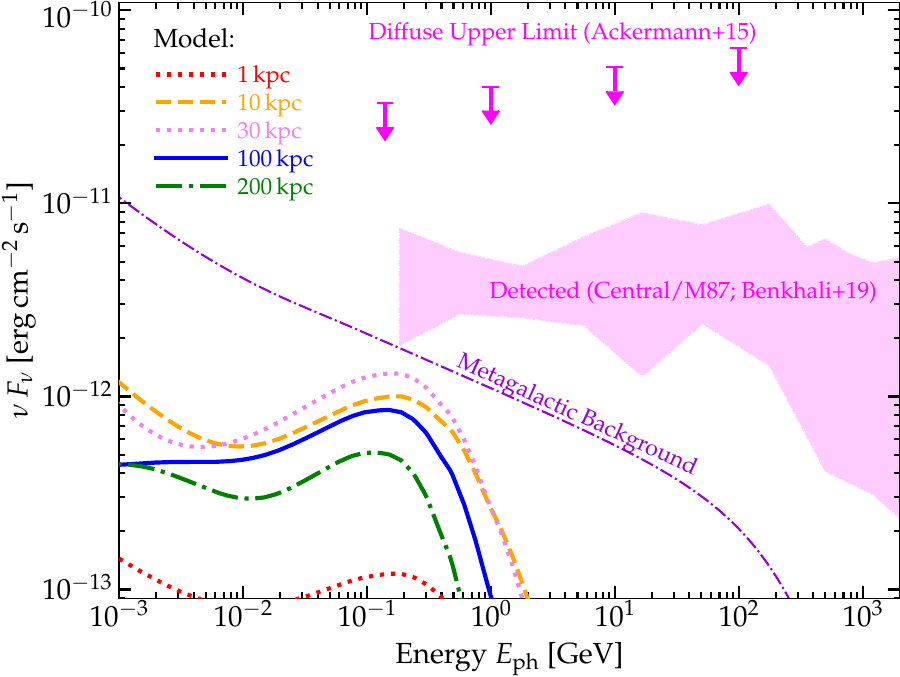}
	\vspace{-0.1cm}
	\caption{Other spectral constraints. 
	{\em Top:} High-frequency radio through infrared, optical, to extreme UV. We show the predicted surface brightness of direct CR emission from the same models in Figs.~\ref{fig:crspec}-\ref{fig:xr}, at different radii, and the observed brightness of the metagalactic background, M87 itself averaged within $<10\,$kpc, and the intracluster light at $10$ and $100\,$kpc. We also show where the observed Galactic column to M87 strongly attenuates the observations. 
	At these frequencies, CR emission is undetectable.
	{\em Middle:} Hard X-rays. We show the backgrounds and diffuse CR emission fluxes (primarily IC of harder M87 IR/optical) integrated in various annuli, plus detections within corresponding radii from NuStar, INTEGRAL, \&\ Suzaku; BAT upper limits; and ASCA+Chandra at $<10\,$keV for reference (see Fig.~\ref{fig:xr}).
	{\em Bottom:} $\gamma$-rays. The emission here is primarily bremstrahhlung at $\sim 10-100\,$MeV, below galactic backgrounds and the detected compact M87$^{\ast}$ source, as well as diffuse upper limits on the total flux outside the core to $\sim$\,degree scales from Fermi (shown).
	\vspace{-0.5cm}
	\label{fig:other.spec}}
\end{figure}

\subsection{Other Wavelengths: UVOIR, Hard X-rays, \&\ Gamma-Rays}
\label{sec:hx.gamma}

Fig.~\ref{fig:other.spec} shows the predicted spectra from CR emission at $10^{10}\,$Hz (3\,cm) through $10^{16}\,$Hz (41\,eV or 300\,\AA). As expected, the direct CR emission is orders-of-magnitude below the metagalactic backgrounds \citep{cooray:2016.extragalactic.background.light.compilations.review,khaire:2019.extragalactic.background.light.spectra} and stellar+dust+gas emission spectra at these wavelengths (compiled from NED averaged within an effective aperture of $10\,$kpc, with the ICL SED specifically coming from the differential/narrow annular regions at $10$ and $100\,$kpc, using the bands and colors available from \citealt{jk:profiles,rudick:2010.virgo.icl.colors,krick:2011.virgo.icl.ir} spanning $4.5\,{\rm \mu m}$ to $U$-band). We also denote in the EUV where Galactic absorption severely limits any constraints, for the standard Galactic column $\sim 4-5\times10^{20}\,{\rm cm^{-2}}$ in the direction of the Virgo core.

More interesting, Fig.~\ref{fig:other.spec} also shows the harder X-rays up to $\sim 400\,$keV. At energies $\gtrsim 10\,$keV, detection of the emission around M87 is challenging, as (1) the relatively low luminosity and extended size on the sky make it comparable to the metagalactic background and often well below the instrument background of hard X-ray detectors like NuSTAR or Switft-BAT, and (2) there is a nearby Seyfert galaxy nucleus NGC 4388 which is extremely bright at these energies, so must be separated (but this is especially difficult given typical angular resolution at these energies). Nonetheless, with higher resolution, NuSTAR has a robust detection within a $\sim 30^{\prime\prime}$ ($<2.4\,$kpc) radius aperture at both $20-40$ and $10-40$\,keV \citep{wong:2017.m87.nustar.detection.agn,imazawa:2021.m87.hardxray.nustar}, where we take the flux range to be the baseline where they detect no evidence for spatial or time variation over baselines as long as years, indicating this comes from extended diffuse gas or jets (and not the AGN nucleus). \citet{abdo:2009.fermi.detection.m87.plus.attempt.to.put.together.other.sed.points} attempt to place an upper limit on emission within a few kpc from Swift-BAT, which we include but stress is highly-sensitive to the assumed instrument background subtraction and NGC4388 separation (likely explaining its apparent tension with the subsequent higher-resolution NuSTAR detections). \citet{walter:2009.integral.detection.m87} detect M87 well above-background with robust separation from NGC 4388, in what appears to be extended emission from $\sim 10-30\,$kpc at $20-60\,$keV.  \citet{dejong:2015.integral.suzaku.hxt.detections.m87} argue that with different background models and multi-year time-averaging this INTEGRAL detection decreases in magnitude but they instead detect extended (time-steady/non-varying) emission from a larger $\sim 100\,$kpc aperture with Suzaku HXD. We show all of these, for completeness. 
Note we include $\sim 0.1-10\,$keV emission in our soft X-ray profiles (Fig.~\ref{fig:xr}), but show it here as well for a couple radii as a useful anchor point, since this is much better measured with instruments like Chandra and ASCA \citep[we specifically show the X-ray spectra directly measured to $10\,$keV in at a matched $\sim 30\,$kpc aperture in][]{allen:2000.hardxr.10keV.virgo.emission.strong.in.central.10s.of.kpc}. 
In any case we see the models are consistent with all of these. Outside the M87 injection region ($\gtrsim 10\,$kpc) they generally predict thermal-like spectra consistent with the soft X-rays which fall off rapidly to $\sim 20\,$keV, with some low-level flux (below the metagalactic background) from CR-IC of higher-energy (infrared-optical) seed photons. 

The diffuse $\gamma$-ray flux is dominated by relativistic bremsstrahlung at $\sim 100\,$MeV. However, Fig.~\ref{fig:other.spec} shows this is quite weak: the extended emission is below the metagalactic background, significantly below (especially at $\gtrsim$GeV energies) the detected $\gamma$-rays from M87$^{\ast}$ and its jets (compiled from \citealt{abdo:2009.fermi.detection.m87.plus.attempt.to.put.together.other.sed.points,hess:2024.m87.core.highenergy.gamma.rays,benkhali:2019.m87.gamma.ray.variability.and.compact.emission}) from Fermi, HESS, MAGIC, and VERITAS, and orders-of-magnitude below the best present upper limits on diffuse $\gamma$-ray emission from Fermi (constraint shown from \citealt{ackermann:2015.fermi.clusters.gamma.ray.upper.limits.diffuse.emission}, but see also \citealt{han:2012.diffuse.cluster.gamma.rays.consistent.with.extended.cosmic.rays.but.highly.dependent.on.point.source.subtraction.dominating.signal.point.source.like.central.contrib.clearly.detected} who obtain very similar upper limits). Because the cutoff is extremely steep, there is basically no predicted emission at TeV for any higher-energy instrument. If we introduced a population of CR hadrons with comparable escaping energy to the leptons at $\gtrsim 1\,$kpc, like in \paperone, this would introduce a second bump from pionic interactions of comparable flux, at $\sim 1-10\,$GeV, with a shallower power-law tail, but it would still be below the central detection. Thus hadron-to-lepton total energy ratios as large as $\sim 30$ (larger than observed in the LISM, once accounting for the integral over the entire energy range; see \citealt{bisschoff:2019.lism.cr.spectra}) could be accommodated in the diffuse gas in Virgo without violating $\gamma$-ray constraints. Of course, if there were such a large hadron population, the total CR acceleration rate of M87 would need to be correspondingly higher, and the CR pressure would be correspondingly larger (so Virgo would be CR-pressure dominated out to a larger radius). Like with Perseus, another allowed scenario is that the leptons invoked here are primarily produced via pionic interactions of hadrons on small scales (i.e.\ the acceleration is hadronic, but those convert an order-one fraction of their energy to leptons before escaping to $\gg$\,kpc scales). We are agnostic to all of these because it is really the leptons which influence all the other observables we consider.\footnote{Note that the models here clearly are well below the $\gamma$-ray upper limits, despite CR pressure dominating the total pressure budget interior to $\lesssim 10-20\,$kpc.  \citet{hess:2023.claimed.upper.limit.cr.pressure.inner.virgo.from.gamma.rays.incorrect.assumptions.all.hadronic.powerlaw.no.curvature.completely.trapped.etc} claimed that the total $\gamma$-ray flux from Virgo sets an upper limit of a few percent of the pressure from CRs, but their constraint (1) applies to an average over much larger radii (does not constrain the CR pressure in the Virgo center); (2) assumes all the CRs are hadronic with negligible leptons; (3) assumes an order-of-magnitude higher gas density $n_{X}$ so is order-of-magnitude weaker when correcting for the CR-IC corrected true gas density in Fig.~\ref{fig:xr}; and (4) assumes an extremely hard, pure-power-law CR spectrum, so a large fraction of the CR energy is at energies $\gtrsim$\,TeV, completely unlike the spectra we model. As such there is no tension between even $10-30\,$times larger CR pressure than we predict, and the $\gamma$-ray observations.}

\subsection{Independent Constraints}
\label{sec:other}

Fig.~\ref{fig:alt} compares various additional, indirect model constraints, on magnetic field strengths, mass/potential profiles, and the Sunyaev-Zeldovich (SZ) effect.

\subsubsection{Magnetic Fields \&\ Tension Intrinsic to the ``Pure Thermal X-rays'' Model}
\label{sec:B}

\begin{figure}
	\centering
	\includegraphics[width=0.97\columnwidth]{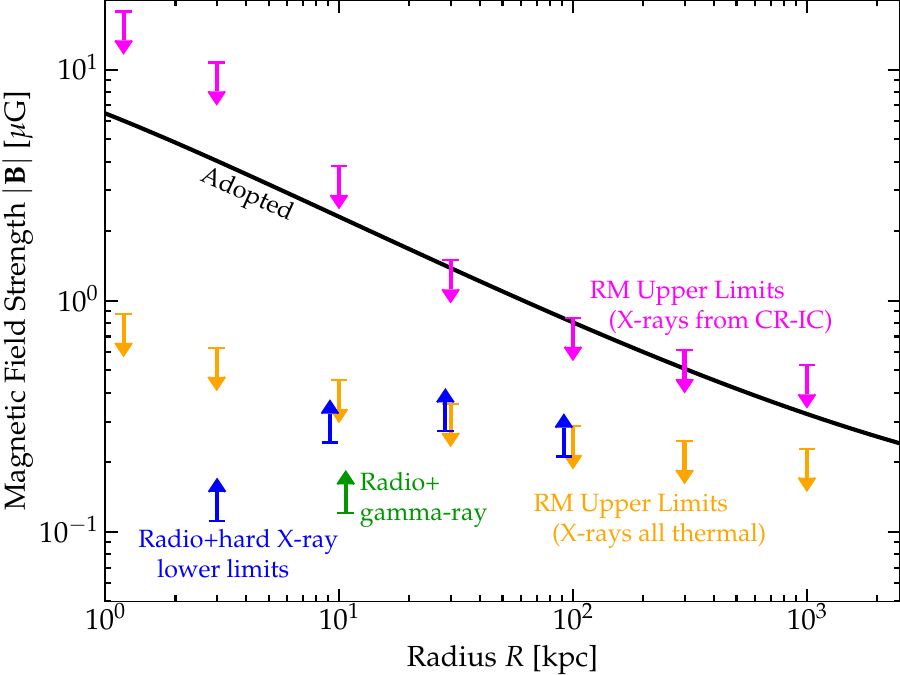}
	\includegraphics[width=0.97\columnwidth]{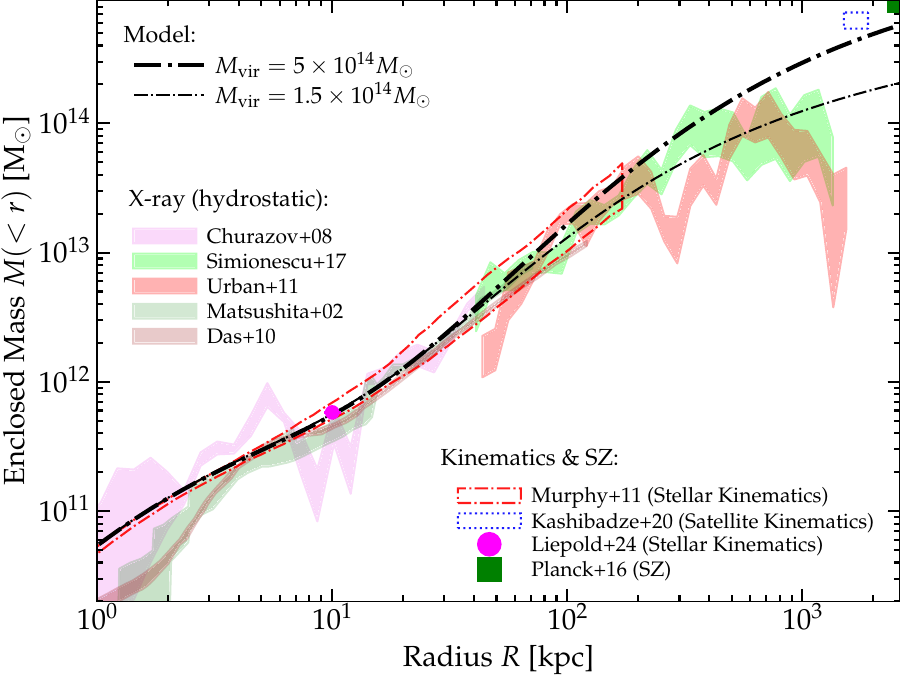}
	\includegraphics[width=0.97\columnwidth]{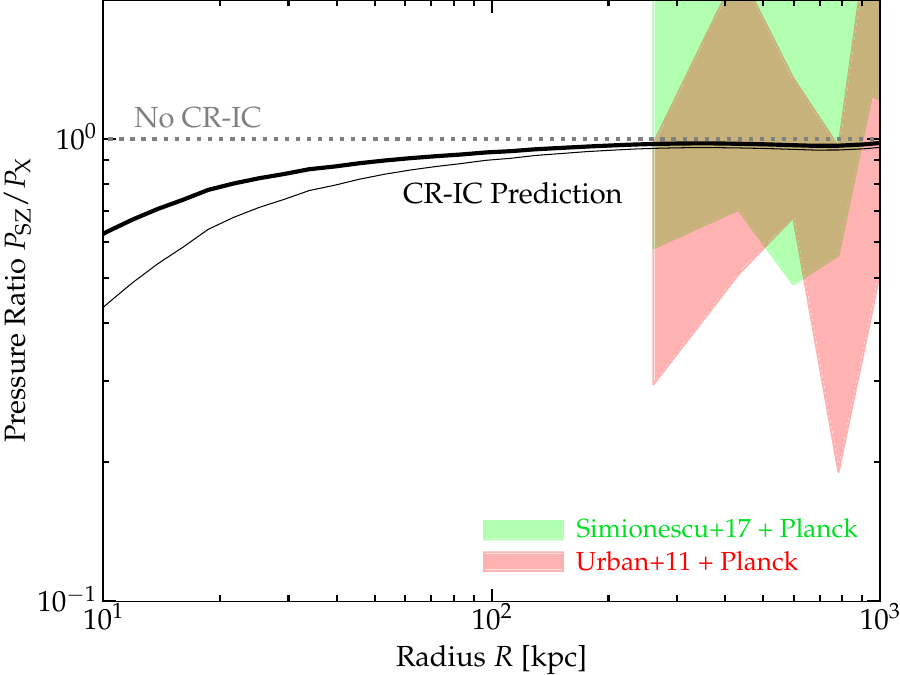}
	\vspace{-0.1cm}
	\caption{Additional constraints versus radius.
	{\em Top:} Model volume-weighted mean diffuse-gas magnetic field $B$, compared to Faraday rotation (RM) constraints. These self-consistently adopt the ``true'' (lower) gas densities of the model in Fig.~\ref{fig:xr} -- in the pure-thermal interpretation of the X-rays, the densities are higher so the RM limits push the upper limits on $B$ to the much lower ($<{\rm \mu G}$ even at $\ll 10\,$kpc) values shown.
	{\em Middle:} Mass profile constraints from X-ray hydrostatic equilibrium modeling versus independent kinematic constraints (stellar/globular/satellite kinematics and SZ). We show two models for the assumed true mass profile (a ``heavy'' and ``light'' Virgo); these differ only at large radii and have no effect on the conclusions here. CR-IC produces deviations at small radii, but these are consistent with the deviations observed. 
	{\em Bottom:} Ratio of thermal pressure inferred from SZ ($P_{\rm SZ}$) and soft X-ray ($P_{X}$; assuming all X-ray is thermal emission) measurements, versus radius. Because of the relatively weak CR-IC contribution and offsetting effects of the temperature and density biases, the effect for a WCC like Virgo is small (compared to SCCs). In Virgo constraints only exist at large radii and are not particularly strong.
	\vspace{-0.5cm}
	\label{fig:alt}}
\end{figure}

There are rich measurements of Faraday rotation measures (RMs) towards the center of M87, from the compact radio core on scales $\lesssim$\,pc down to near-horizon (au) scales \citep{matveyenko:2014.m87.rm.structure.spatial.gradients,kuo:2014.rms.m87.core.sma.all.from.core.gas,li:2016.modeling.m87.3c84.rms.jet.vs.adaf,avachat:2016.m87.jet.rm.structure.variability.rms.from.within.jet,algaba:2016.m87.jet.rm.spatially.resolved.kpc.scales.coming.from.within.or.very.close.to.jet.much.smaller.subkpc,pasetto:2021.m87.jet.helical.field.rms,tseunetoe:2022.m87.detailed.rm.modeling.jet.sheaths.and.bases.not.larger.icm,peng:2024.m87.jet.structure.tiny.scale.structure.rms.inside.jet.and.very.close.to.smbh}. As the number and quality of RM measurements have increased, a very robust result is that the RMs vary dramatically on small angular/spatial ($\lesssim$\,pc) and time ($\lesssim$\,yr) scales, which places very strong upper limits on the RM contribution from the diffuse ICM with fields on $\gtrsim$\,kpc coherence scales. \citet{park:2019.m87.rms.detailed.modeling.huge.spatial.temporal.datasets.strong.upper.limits.outside.bondi.radius} consider a large ensemble of these and argue this is approximately $|{\rm RM}_{\rm ICM}| \lesssim 130\,{\rm rad\,m^{-2}}$, with about a factor of $< 2$ uncertainty. This is consistent with \citealt{algaba:2016.m87.jet.rm.spatially.resolved.kpc.scales.coming.from.within.or.very.close.to.jet.much.smaller.subkpc,tseunetoe:2022.m87.detailed.rm.modeling.jet.sheaths.and.bases.not.larger.icm,peng:2024.m87.jet.structure.tiny.scale.structure.rms.inside.jet.and.very.close.to.smbh}, who all place upper limits $\ll 1000\,{\rm rad\,m^{-2}}$ from the ICM, and consistent with the qualitative conclusions of the other RM studies above that all argue the RMs must come from within or very close to the inner accretion disk/jet core, jets, or jet sheaths. Given ${\rm RM} = 812\,{\rm rad\,m^{-2}}\,\int_{0}^{\infty} (n_{e}/{\rm cm^{-3}})\,(B_{\|}/{\rm \mu G})\,(d\ell/{\rm kpc})$, this implies an upper limit (assuming isotropic fields) of $|{\bf B}| \lesssim B_{\rm max} \sim 0.28\,{\rm \mu G} / \int_{0}^{\infty}  n_{e} d\ell / ({\rm cm^{-3} kpc} )$. For a given model of $n_{e}(r)$, we therefore immediately obtain the corresponding $B_{\rm max}$, shown in Fig.~\ref{fig:alt}. We plot this first for the default model here, where $n_{e}$ is relatively low and the X-rays come primarily from CR-IC, and show it is compatible with the magnetic field profile we assume. Note that we specifically adopt a model $B(r)$ near the high-end allowed, because any \textit{weaker} field means that it is even more likely that for the given observed radio emission, the X-rays should primarily come from CR-IC, so this is the ``worst-case scenario'' for the CR-IC models here. But we also show $B_{\rm max}(r)$ in the traditional interpretation of the soft X-rays as thermal emission: because producing the observed SB with just thermal emission requires much larger densities $n_{e}$ at a given $r$, the RM upper limit then requires $B_{\rm max}$ is much lower, $\ll {\rm \mu G}$ at essentially all radii even in the M87 ISM, which seems unphysical. 

% owen:1990.original.m87.rm.measurements -- old M87 RM data, first detections, low-res, interpreted as large ~kpc screen, still just central kpc though
% capetti:1997.m87.jet.polarization.rms.along.jet - early ish measurements, but already showing with higher-res much more variability in space along the jet

Note that we can also estimate an upper limit on $B$ from kinematics, arguing that there cannot be more than a certain fraction of additional pressure or it would unbind the cluster. We have computed this, but these constraints are significantly weaker than the RM constraints shown. 

We can also place a \textit{lower} limit on $B$, $|{\bf B}| \gtrsim B_{\rm min}$, from the combination of radio profiles and hard X-ray emission. This is a standard technique \citep[e.g.][]{cova:2019.cluster.ic.upper.limits.B.lower.limits,creech:2024.nustar.perseus.obs}, and uses the fact that CR leptons which radiate at a synchrotron frequency $\sim 10\,{\rm MHz}\,B_{\rm \mu G}\,E_{\rm GeV}^{2}$ will also produce CR-IC from the CMB at X-ray energy $\sim 3\,{\rm keV}\,E_{\rm GeV}^{2}$, with a ratio of specific emissivity $\propto e_{\rm cmb}/e_{\rm B}$. Given the ranges allowed, the most powerful constraint comes from combining the 20\,MHz radio profile with the $\sim 20-60\,$keV X-ray emission (for which we take the highest values in Fig.~\ref{fig:other.spec}, as any lower value would lead to an even-larger upper limit in $B_{\rm min}$, so this is the most conservative possible choice). We plot this over the range where there is some constraint on the integrated flux in both bands. We can also estimate the same by combining the GHz radio with GeV $\gamma$ ray emission (shown), but this is somewhat less constraining. In either case this sets lower limits around $\sim 0.2-0.3\,{\rm \mu G}$. 

We immediately see that there is some tension, purely empirically, between the ``thermal-only'' assumption for soft X-rays coupled to the observed RMs and the observed radio and hard X-ray upper limits. This is not yet at a level that we can say the thermal-only soft X-ray models are conclusively ruled out from Fig.~\ref{fig:alt} alone, but the tension is noteworthy.

\subsubsection{Mass Models and Evidence for Non-Thermal Components}
\label{sec:mass}

Fig.~\ref{fig:alt} shows the 3D mass profile $M(<r)$ of Virgo, from different reconstruction methods. We compile several non-X-ray studies using stellar plus globular cluster kinematics \citep{murphy:2011.new.m87.models,liepold:2023.m87.dynamical.masses}, satellite kinematics \citep{kashibadze:2020.virgo.galaxy.number.density.profile.and.cluster.kinematics}, and SZ \citep{planck:2016.sz.virgo.signal.only.large.radii.constrained.but.those.close.to.xray}. In the latter, the SZ is not used for a pressure profile, but rather, to infer the baryonic mass on the assumption of the X-ray fitted temperature (at large radii) being correct, then assuming a universal baryon fraction. We compare several X-ray studies assuming hydrostatic equilibrium \citep{matsushita:2002.m87.virgo.cluster.obs.metallicity.drop.temperature.fitting.challenges,churazov:2008.virgo.mass.profile.kinematics.vs.xrays.as.measurement.of.nonthermal.pressure,das:2010.m87.xray.hydrostatic.mass.profiles,simionescu:2017.virgo.profiles.outskirts}, and our own reconstruction using the de-projected X-ray profiles in \citet{urban:2011.virgo.xray.profiles.to.rvir}. We compare the fiducial adopted mass model ($M_{\rm vir} = 6 \times10^{14}\,{\rm M_{\odot}}$, $M_{200c} \approx 5 \times10^{14}\,{\rm M_{\odot}}$) and a ``light'' halo $M_{\rm vir} = 1.5 \times10^{14}\,{\rm M_{\odot}}$ ($M_{200c} \approx 1.2 \times10^{14}\,{\rm M_{\odot}}$). Recall, the CR-IC+thermal emission models here reproduce the observed X-ray profiles in their ``apparent'' X-ray emission, so the inferred mass profile from them will necessarily be the same as the observations plotted.

%\citet{kashibadze:2020.virgo.galaxy.number.density.profile.and.cluster.kinematics} - number density profile of galaxies vs cluster-centric distance, but also dispersions, etc, but no mass model (but can use for 'giant halo' model - all at >200 kpc though)
% lee:2015.filament.infall.virgo.very.large.virial.mass - filament infall rate estimate of virial mass of virgo from very large radii, gets quite large = 8e14

A few interesting points emerge. The X-ray and kinematic data differ significantly at many radii, contrary to earlier claims in e.g.\ \citet{churazov:2008.virgo.mass.profile.kinematics.vs.xrays.as.measurement.of.nonthermal.pressure} which featured much larger systematic uncertainties in both kinematic models (e.g.\ their reference \citealt{wu.tremaine:2006.globular.m87.mass.distribution}, which quoted factor $\sim 6$ uncertainties at the salient radii) and X-ray data (the large radial fluctuations in Fig.~\ref{fig:alt} in the \citealt{churazov:2008.virgo.mass.profile.kinematics.vs.xrays.as.measurement.of.nonthermal.pressure} data). 
Most notably, at small radii $\lesssim 20\,$kpc, the updated X-ray data and hydrostatic masses with much smaller error bars from the same authors in \citet{das:2010.m87.xray.hydrostatic.mass.profiles} (also \citealt{matsushita:2002.m87.virgo.cluster.obs.metallicity.drop.temperature.fitting.challenges}) appear to under-estimate the enclosed mass by factors of several (while \citealt{churazov:2008.virgo.mass.profile.kinematics.vs.xrays.as.measurement.of.nonthermal.pressure} oscillates by factors of several around $M(<r)$), when compared to the newer kinematic constraints (e.g.\ \citealt{murphy:2011.new.m87.models,agnello:2014.m87.gc.kinematics.updated.mass.models,liepold:2023.m87.dynamical.masses}), that leverage newer, higher-resolution resolved HST+IFU spectroscopy, expanded globular kinematics, gas kinematics, and more sophisticated triaxial jeans modeling with expanded black hole+stellar+dark matter models. 
At large radii $\gtrsim 50\,$kpc the same factor-of-few oscillations are evident, and the X-ray data appears to systematically favor a surprisingly low total mass of $\sim 1\times 10^{14}\,{\rm M_{\odot}}$ \citep{urban:2011.virgo.xray.profiles.to.rvir,simionescu:2017.virgo.profiles.outskirts}, in apparent contradiction to constraints from (1) SZ \citep{planck:2016.sz.virgo.signal.only.large.radii.constrained.but.those.close.to.xray}, (2) satellite kinematics \citep{kashibadze:2020.virgo.galaxy.number.density.profile.and.cluster.kinematics}, and (3) large-scale infall motions of filaments and galaxies outside of the Virgo virial radius \citep{lee:2015.filament.infall.virgo.very.large.virial.mass}, all of which appear to favor a virial mass $\sim 5-8\times10^{14}\,M_{\odot}$.\footnote{Also, regardless of the ``light'' or ``heavy'' halo, both require somewhat high values of the concentration parameter $c_{\rm vir} \equiv R_{\rm vir}  / R_{s,\,{\rm NFW}}$, specifically $c_{\rm vir} \sim 7.5$ for the ``heavy'' halo and $c_{\rm vir} \sim 10$ for the ``light'' halo ($c_{200c} \sim 5.6,\,7.5$, respectively), to match the observations at $\sim 40-150\,$kpc -- about a $+2\sigma$ deviation from the standard $c-M$ relation \citep{merten:2015.clash.cluster.concentration.mass,biviano:2017.cluster.concentration.mass.collection}.} 
We stress that this has no effect on our modeling in this paper, as it only becomes important ourside $\gtrsim 200-300\,$kpc where CR-IC is negligible, and CRs cannot plausibly explain the discrepancy, but it suggests (as noted in several of the papers above) that either some of the observations are incorrect or that there is substantial non-equilibrium (and/or non-hydrostatic, or non-spherical) structure in Virgo at $\gtrsim 0.2-0.3\,R_{500}$. There are many possibilities.
The discrepancy at smaller radii is more interesting for the models here. Note that even where the X-ray and kinematics appear to agree well (e.g.\ $\sim 15-40\,$kpc), the rms deviations between the two are $\sim 20-50\%$, comparable to the systematic uncertainties in both (just measuring the X-ray profile along different azimuthal angles, or using slightly different tracer populations for the stellar kinematics, give different results at this level). And at small radii $\lesssim 10\,$kpc, where the kinematic constraints are best, there is a much larger factor-of-few discrepancy, and this has been shown and discussed in many previous studies as the kinematic data have improved (and similar discrepancies have been noted for the locally-peaked X-ray profiles around other, individual galaxies within Virgo, e.g.\ \citealt{shen:2010.virgo.galaxy.very.massive.bh.and.xray.kinematic.data.disagree.strongly} and references therein). This makes it easy, for example, to accommodate a predominant CR pressure, buoyant motions in the ICM, and non-equilibrium effects.

\subsubsection{The Sunyaev-Zeldovich Effect: Less Sensitive in Weak Cool Cores}
\label{sec:SZ}

Fig.~\ref{fig:alt} also shows the predicted ratio of SZ-inferred pressure $P_{\rm SZ}$ to X-ray inferred pressure $P_{X}$, on the assumption that the observed ``apparent'' X-ray pressure (Fig.~\ref{fig:xr}) is interpreted as thermal, together with the same ratio from Planck observations in the outer ICM \citep{planck:2016.sz.virgo.signal.only.large.radii.constrained.but.those.close.to.xray}. As described in \citet{hopkins:2025.cr.ic.clusters.ideapaper}, because of CR-IC boosting $n_{X}$, the X-ray pressure $P_{X}$ is over-estimated in the central regions so $P_{\rm SZ}/P_{X}$ is suppressed, but this is a mild effect for a weak CC like Virgo, compared to SCCs discussed therein and in \paperone. We see that the predictions are completely compatible with present SZ constraints, but these are only at very large radii. Given Virgo's large size on the sky, potential contamination by the jets at $R \lesssim 10\,$kpc, and the relatively small magnitude of the predicted effect, it is unlikely the predicted difference will be robustly observed with present instruments.

\section{Differences from SCCs like Perseus?}
\label{sec:sccs}

Here we briefly comment on the major differences between plausible models like those here for CR-IC in a WCC like Virgo, versus CR-IC models like those in \paperone\ for SCCs like Perseus. First, per \S~\ref{sec:cf}, a key physical interpretation difference is that in SCCs like Perseus, the CR-IC models still favor (and probably require) \textit{some} cooling and multi-phase gas in the ICM, just 100$\times$ less than the naive classical SCC cooling-flow problem would imply. Thus an ``apparent'' SCC can become a ``true'' WCC. Starting from an apparent WCC like Virgo, where there is no evidence for cooling gas at $\gtrsim $\,kpc scales, the apparent WCC can become a ``true'' NCC.

Of course, some properties are different in the models for Virgo, but in a way that is not interesting for the physics we explore: the CR injection rate $\dot{E}_{\rm cr}$ is $\sim 100$ times lower in our Virgo model compared to Perseus, but this simply reflects the observed difference in the radio+$\gamma$-ray+X-ray luminosities of the clusters. And of course the large-radii/``underlying'' gas density/temperature/metallicity profiles differ but just as expected for different mass clusters (as the cluster masses differ by an order of magnitude). 

\subsection{Softer Injection Spectra}
\label{sec:sccs:injection}

Turning to non-trivial differences relevant for the actual physics here, first we note that comparing to \paperone\ and Perseus, the nominally favored CR-IC models here have somewhat softer effective CR injection (emergent from $\sim$\,kpc) spectra. As discussed in \S~\ref{sec:TX}, some of this is driven by our arbitrary assumption that the ``true'' temperature $T$ is constant at small radii, so the apparent lower $T_{X}\sim 1\,$keV must be driven by CRs, which therefore requires a soft CR spectrum at small radii. But as emphasized there, this is not necessary -- especially in CR-pressure dominated central regions, the physical temperatures could and perhaps should be cooler, so (because the effect of CR-IC on the inferred $T_{X}$ is relatively weak), modeling that would allow for significantly harder injection spectra. However there is still a clear indication of softer injection spectra from both radio and hard X-ray observations: while Perseus is clearly detected by NuStar, Hitomi, Swift-BAT, Integral, and many others with a very significant hard X-ray component, per \S~\ref{sec:hx.gamma} that is weaker in Virgo. More clearly, the radio spectral index in the central regions is significantly softer in Virgo: in the central ($\sim$\,kpc, here, not the radio core on sub-pc scales) regions from LOFAR observations and other low-frequency ($\lesssim 150\,$MHz) data, the spectral index never gets much shallower than $\alpha \sim 0.6$ or so (outside of very specific ``hot spots'' like the jet shocks/lobes/``ears''), while for the central regions of NGC 1275/3C 84 (the Perseus BCG) this index approaches more like $\alpha \sim 0$. So this is not a difference unique to the CR-IC interpretation, but a purely empirical question. 

In terms of explanations, it is plausible to understand if, for example, the acceleration region in the M87 core is more compact (compared to the more extended jets/lobes/cocoons/shocks around NGC 1275), and contained within the dense gas, strong magnetic field, and strong radiation energy-density region on sub-kpc scales around M87$^{\ast}$. The emergent spectrum at $\gtrsim$\,kpc could therefore plausibly be more significantly influenced by losses ``on the way out.'' But there are many other possibilities as well.

\subsection{Transport Parameters} 
\label{sec:sccs:transport}

Looking at the CR transport parameters $\kappa$ (effective isotropic diffusivity) and $v_{\rm st}$ (effective bulk streaming speed, including advection/super and \Alf{ic} streaming, super-diffusion, or radially-dependent diffusion), although the value of the total ``effective transport speed'' $v_{\rm eff} \sim v_{\rm st} + \kappa / r$ is broadly similar to that in \paperone\ from $\sim 10-100\,$kpc, the fiducial values here involve a lower diffusivity $\kappa$ and higher streaming speed $v_{\rm st}$, compared to the fiducial Perseus model in \paperone. This preference is driven entirely by the shape of the central radio and X-ray surface brightness profiles in Virgo versus Perseus: Virgo has a much steeper radio/X-ray brightness and therefore inferred-density $n_{X}$ profile at all radii $\lesssim 100\,$kpc, compared to Perseus (e.g.\ $\epsilon_{X} \propto n_{X}^{2} \propto r^{-2}$ versus $\propto r^{-1}$, a difference of a factor of $\sim 100$ in emissivity over this radial range). Given the simplicity of our default models, where the injection is point-like, both $\kappa$ and $v_{\rm st}$ are constant, and the clusters are spherically symmetric and in steady-state, the steady-state emissivity from CR-IC scales $\propto e_{\rm cr} \propto 1/(v_{\rm st}\,r^{2})$ in the streaming-dominated limit ($v_{\rm st} \gg \kappa/r$) or $\propto 1/(\kappa\,r)$ in the diffusion-dominated limit. 

This could arise, in principle, from variations in the microphysics of CR scattering and different plasma and/or CR properties (values of $B$, $n$, $T$, strength of turbulence, CR energy density, etc.). What actually sets the microphysical CR scattering rates on $\sim 0.1\,{\rm au}$ scales (the spatial scales relevant for magnetic field fluctuations which scatter CRs of the energy range of interest here), which in turn controls the physical values and scalings of $\kappa$ or $v_{\rm st}$, is deeply physically uncertain (even in the LISM where CR measurements are most accurate), and as a result different first-principles plasma physics models predict qualitatively different scalings of these parameters that can differ by as much as $\sim 10$ orders-of-magnitude in the CGM/ICM \citep{hopkins:2020.cr.transport.model.fx.galform,hopkins:cr.transport.constraints.from.galaxies,hopkins:2021.sc.et.models.incompatible.obs,ponnada:2023.fire.synchrotron.profiles,ponnada:2023.synch.signatures.of.cr.transport.models.fire,fitzaxen:2024.cr.transport.into.gmcs.suppressed.starforge,ponnada:2024.fire.fir.radio.from.crs.constraints.on.outliers.and.transport,su:2025.crs.at.shock.fronts.from.jets.injection,lu:2025.cr.transport.models.vs.uv.xray.obs.w.cric}. So it is ambiguous to interpret this in terms of any specific CR microphysics. 

Alternatively, this effect could be related to other, more resolvable properties. If in Virgo, advection in buoyant modes or winds or the jets themselves is more important, the CR transport could be more ``streaming-like'' because CRs are simply carried by bulk motions. Alternatively, as noted in \S~\ref{sec:sccs:injection}, this could also relate to the effective spatial scale of injection. The flatter profile in Perseus could arise not because of more-diffusive transport from a point-source-like injection, but rather from the same transport parameters as Virgo, but a more spread-out injection region extending to several tens of kpc (the size of the lobes/cavities). Future modeling in more detailed cluster simulations with CR dynamics can make more robust predictions for these, but in simplified analytic models like those here these different scenarios are largely degenerate.

\subsection{Lack of a Giant Radio Halo}
\label{sec:sccs:giant}

In \paperone, we showed that the CR-IC models can also produce a giant radio halo extending to $\sim$\,Mpc. This does not come from CRs injected by the central source, no matter how luminous, as the leptons cannot reach $\gtrsim$\,Mpc scales before losing their energy to IC. But per \citet{hopkins:2025.crs.inverse.compton.cgm.explain.erosita.soft.xray.halos,hopkins:2025.cr.ic.clusters.ideapaper}, every cluster galaxy, even low-mass (Milky Way-like or smaller) galaxies, produces its own CR ``mini-mini-halo'', which appears in IC around isolated Milky Way and M31-mass galaxies in ROSAT and eROSITA soft X-ray survey stacks \citep{zhang:2024.hot.cgm.around.lstar.galaxies.xray.surface.brightness.profiles}, and is clearly evident in the extended $\gamma$-ray emission observed in the Milky Way \citep{acero:2016.gamma.ray.constraints.cr.emissivity,yang.2016:diffuse.gamma.ray.cr.profile.constraints,pothast:2018.radial.mw.gamma.profile.modeling.fermi} and M31 \citep{abdo:2010.outer.mw.gamma.ray.emission,karwin:2019.fermi.m31.outer.halo.detection}. Each extends to $\sim 100\,$kpc, but their individual surface brightness is extremely low (lower than the central-galaxy halo by approximately the ratio of their CR injection rates, and further suppressed by weaker magnetic fields in the cluster outskirts), but because they effectively overlap in projection, they collectively can be observed in a case like Perseus (\paperone). The same process should happen in Virgo, and we in fact calculate this collective contribution in the profiles shown in Fig.~\ref{fig:radio} using the Virgo projected galaxy number density compiled in \citet{kashibadze:2020.virgo.galaxy.number.density.profile.and.cluster.kinematics} and assuming a total CR injection rate from Virgo satellites a factor of $\sim 0.1$ times the M87 injection rate (the same ratio observed in Perseus, and consistent with the ratio of radio luminosity of the satellites, though that could be as low as $\sim 0.02-0.1$ \citealt{capetti:2009.radio.virgo.survey,vollmer:2013.radio.virgo.survey,spasic:2024.radio.virgo.survey}). But this immediately explains why the predicted giant halo is negligible in Fig.~\ref{fig:radio}: (1) the Virgo injection rate is $\sim 100$ times smaller than Perseus, and (2) Virgo's extended halo is $\sim 100$ times larger on the sky. This not only makes it infeasible to observe the whole cluster, but means the predicted surface brightness of the giant halo from this source is $\sim 10^{4}$ times fainter. Given that the giant halo in Perseus has only \textit{just} been detected by new, extremely deep integrations from the most sensitive instruments and wavelengths in LOFAR \citep{vanweeren:2024.perseus.giant.radio.halo.filled.electrons.just.tiny.fraction.high.energy}, it is clear that we do not expect to see the same in Virgo.

\section{Conclusions}
\label{sec:conclusions}

We show that a simple model of CR leptons injected in M87$^{\ast}$ with a kinetic luminosity comparable to the usual value inferred from the compact radio and $\gamma$-ray spectra/variability, and propagating outwards in Virgo in steady-state with a streaming speed $\sim 100\,{\rm km\,s^{-1}}$ (subject to standard losses), can naturally explain a wide variety of observations from radio ($\sim 10^{6}-10^{11}\,$Hz) and soft X-rays ($\sim 0.1-10\,$keV), without contradicting any limits from harder X-ray or $\gamma$-rays, SZ, Faraday rotation, or dynamical/mass/potential reconstruction constraints, at radii $\sim 1-1000\,$kpc. 
These models predict that the central, steeply-rising soft X-ray surface brightness interior to $\lesssim 20\,$kpc has a significant contribution from thermal-continuum-like (highly curved/non-power-law) inverse-Compton scattered CMB photons. 

This naturally explains the ``cooling flow problem'' in M87: the true cooling rate in the model is much lower ($\sim 0.1\,{\rm M_{\odot}\,yr^{-1}}$) than the cooling rate inferred if one assumes the X-rays are purely thermal ($\sim 10\,{\rm M_{\odot}\,yr^{-1}}$). It simultaneously explains the shape of the central CC soft X-ray profiles in surface brightness $S_{X}$, inferred thermodynamic properties ($n_{X}$, $T_{X}$, $P_{X}$, $K_{X}$, $\dot{M}_{X}$), and metallicity $Z_{X}$ (explaining the apparent metallicity suppression inferred in $\sim$\,keV X-rays but not optical/UV or harder/Fe-K X-rays in $Z$ within a few kpc), as well as the apparent tight correlation between the M87 jet/radio power and X-ray cooling luminosity (both powered by the same lepton source). It also explains the radio spectrum and surface brightness profiles of the diffuse, ultra-steep volume-filling radio halo of M87 on these scales, and its clear spectral curvature and steepening with distance from M87$^{\ast}$. 

We further show that new Faraday rotation measure (RM) limits on the magnetic field strongly argue for a significant CR-IC contribution at small radii, as otherwise (with the much larger gas densities required to reproduce the soft X-rays in a thermal-only scenario) the RM upper limits to $B$ in the diffuse cluster gas at $\lesssim 30\,$kpc would be unphysically low ($\lesssim 0.1\,{\rm \mu G}$), and in tension with lower limits on $B$ from the combination of the radio detections and hard X-ray emission upper limits. 

Together, this suggests that at least some apparent ``weak cool core'' (WCC) clusters with central radio galaxies, of which Virgo is often considered prototypical, may in fact have true thermodynamic properties more consistent with a relaxed-but-not significantly cooling cluster (at least outside $\gtrsim 1\,$kpc), plus central CR-IC emission. Even more robustly, these indicate that at $\lesssim 10$\,kpc around M87, CR pressure must be a significant (and likely dominant, at $\lesssim 5\,$kpc) fraction of the total pressure in the diffuse gas.

Our model does not explain every feature of Virgo in radio and X-rays, of course. For example, we assume for simplicity spherical symmetry with point-like central CR injection, but it is clear that CRs are also being accelerated or re-accelerated in some special extended jet termination/shock/``radio ear'' regions related to the extended jets out to few-kpc scales, which are specifically associated with the regions where higher-frequency synchrotron and less-curved radio spectra are observed. Future work using full dynamical CR-MHD simulations, with injection and transport parameters motivated by the simple analytic models here, would be particularly informative. Interestingly, compared to more extreme radio+X-ray sources like Perseus, we do not predict a detectable giant radio halo (owing to the weaker overall lepton injection and much lower satellite radio source density on the sky), and the radio data imply a softer CR spectrum escaping or accelerated outside of the central $\sim$\,kpc, with a somewhat more ``streaming-like'' (or super-diffusive, or $\kappa\propto R$) profile (directly related to the steeper central surface brightness in radio and X-rays observed in Virgo versus Perseus). We discuss a variety of physics that could give rise to these more subtle differences, but given the orders-of-magnitude uncertainty in how CR transport physics scales with ICM/plasma properties, it is difficult to tie these to specific physical models.

\paperone\ reviewed future observational tests of the picture here. Future spatial-and-spectrally-resolved (de-projected) X-ray spectra at $\sim 1\,$keV to measure $Z_{X}$ within the CC ($\lesssim 10\,$kpc) combined with independent optical/UV gas-phase metallicities $Z_{\rm true}$ constrain CR-IC. Optical data already exists at small-$R$ but future X-ray missions like ATHENA could greatly improve the X-ray side. In radio, mapping diffuse synchrotron (in the CC but outside the jets) at $\lesssim 20\,$MHz will further improve the overlap of the energy range where leptons emitting in soft X-rays via CR-IC would also emit synchrotron, to test if their implied spectral shapes and emissivities are consistent with the models here (or if, for example, a significant CR-IC emission in X-rays is avoided by the presence of a sudden, very steep cutoff in the radio spectrum just below $\sim 20-40\,$MHz). Per above, constraints on Faraday rotation in M87 are already powerful, but future independent measurements from background fast radio bursts with impact parameters $\lesssim 10\,$kpc can rule out pure-thermal emission in X-rays (given the observed radio spectrum) by placing RM upper limits below several thousand ${\rm rad\,m^{-2}}$. Higher-resolution Sunyaev-Zeldovich observations probing $\ll 20\,$kpc would be extremely powerful as a constraint on the true thermal pressure, though the ``pressure deficit'' (ratio $P_{\rm SZ}/P_{\rm X}$) is small at most radii outside $\sim 1\,$kpc as shown above, and these are possible in principle (it requires combining instruments to capture large modes on the sky with high-resolution in the center, but the M87 core has a surface brightness at $\sim 140\,$GHz nearly $\sim 1000$ times smaller than NGC 1275 in Perseus, and is larger on the sky, so contamination is much less an issue). Strong $\gamma$-ray constraints require sub-arcminute resolution and at least $\sim 100$ times improved background control (diffuse surface-brightness sensitivity) relative to instruments like Fermi, but this would be powerful at $\sim 10-1000$\,MeV energies. Constraints from ionization of dense cold phases in the ICM as available in Perseus can directly measure the number density of low-energy CRs, but are only possible if a significant mass of dense atomic/molecular gas exists embedded within the Virgo ICM at $\sim 10\,$kpc; thus far searches have only produced upper limits at these radii.

\begin{acknowledgements}
Support for PFH was provided by a Simons Investigator Grant.
\end{acknowledgements}

\bibliographystyle{mn2e}
%\bibliography{/Users/phopkins/Dropbox/Public/ms}
\bibliography{ms_extracted}

\end{document}